\documentclass[11pt,a4paper]{article}
\usepackage{jheppub}
\usepackage{amsmath}
\usepackage{multicol,bbm}
\usepackage{enumerate}
\usepackage{bibentry}
\usepackage[toc,page]{appendix}
\graphicspath{{./Figs/}}

\newcommand{\beqs}{\begin{equation*}}
\def\beq{\begin{equation}}

\newcommand{\eeqs}{\end{equation*}}
\def\eeq{\end{equation}}

\def\beq{\begin{equation}}
\def\eeq{\end{equation}}
\def\be{\begin{equation}}
\def\be{\begin{equation}}
\def\ee{\end{equation}}
\def\ba{\begin{eqnarray}}
\def\ea{\end{eqnarray}}
\def\bea{\begin{eqnarray}}
\def\eea{\end{eqnarray}}
\def\eq{\begin{equation}}
\def\eqe{\end{equation}}
\def\eqa{\begin{eqnarray}}
\def\eqae{\end{eqnarray}}

\def\beqa{\begin{eqnarray}}
\def\eeqa{\end{eqnarray}}

\newcommand{\beqas}{\begin{eqnarray*}}

\newcommand{\eeqas}{\end{eqnarray*}}

\title{Soft theorems from anomalous symmetries}
\author[a]{ Yu-tin Huang}
\author[b]{Congkao Wen}
\affiliation[a]{Department of Physics and Astronomy, National Taiwan University, Taipei 10617, Taiwan, ROC} 
\affiliation[b]{I.N.F.N. Sezione di Roma ``Tor Vergata", Via della Ricerca Scientifica, 00133 Roma, Italy}
\emailAdd{yutinyt@gmail.com,Congkao.Wen@roma2.infn.it} 

\abstract{We discuss constraints imposed by soft limits for effective field theories arising from symmetry breaking. In particular, we consider those associated with anomalous conformal symmetry as well as duality symmetries in supergravity. We verify these soft theorems for the dilaton effective action relevant for the $a$-theorem, as well as the one-loop effective action for $\mathcal{N}=4$ supergravity. Using the universality of leading transcendental coefficients in the $\alpha'$ expansion of string theory amplitudes, we study the matrix elements of operator $R^4$ with half maximal supersymmetry. We construct the non-linear completion of $R^4$ that satisfies both single and double soft theorems up to seven points. This supports the existence of duality invariant completion of $R^4$.}

\begin{document}
\begin{large}

\maketitle 
%%%%%%%%%%%%%%%%%%%%%%%%%%%%%%%%%%%%%%%%%%%%%%%%%%%%%%%%%%%%%
\section{Introduction and motivations}
%%%%%%%%%%%%%%%%%%%%%%%%%%%%%%%%%%%%%%%%%%%%%%%%%%%%%%%%%%%%%
Soft behaviours of scattering amplitudes for Goldstone bosons encode the detailed structures of the underlying spontaneously broken symmetry. This property has played important roles in studying UV properties of extended supergravity theories~\cite{Lance1, ElvangSUSYCT, ElvangSUSYCTE7}, as well as constraining effective actions~\cite{Low1,Jaro1} that are associated with particular broken symmetries, and computing their corresponding scattering amplitudes~\cite{Jaro2}. Here we will be interested in cases where the effective actions also include terms that are associated with anomalies. Prominent examples include the dilaton effective action that encodes the conformal anomaly, relevant for the $a$-theorem~\cite{KS, Elvang6D}, as well as the one-loop effective action for $\mathcal{N}=4$ supergravity, which contains anomalous terms for the U(1) duality symmetry~\cite{N4Anom}.

In practice the presence of anomalies implies that the associated current is no-longer conserved, which leads to a non-vanishing single-soft limit for the goldstone bosons. For the case of dilaton effective action, we will demonstrate that the limit yields:\footnote{Soft theorems for theories with spontaneously broken conformal symmetry were also studied in~\cite{Rutgers} recently.}
\eq
 M_{\alpha,\beta,\varphi}|_{q\rightarrow0}=\frac{2}{d-2}\frac{1}{f^{(d-2)/2}}\sum_{i}\mathfrak{D}_iM_{\alpha,\beta}\,.
\eqe
where $q$ is the momentum of dilaton $\varphi$ which we take to be soft, $\mathfrak{D}_i$ the single site dilatation operator, $d$ the space time dimensions and $f$ the mass scale. Using explicit amplitudes derived from the $d$-dimensional dilaton effective action~\cite{Elvang:2012yc}, we verify that the above result indeed holds. Note that the effective action is defined up to coefficients which reflect the $a$-anomaly, as well as the freedom to add Weyl invariant operators. Our analysis shows that leading soft constraints are satisfied irrespective of these particular coefficients. To obtain non-trivial constraints on $\Delta a$, one must consider other symmetries. As an example, using R-symmetry constraints we can show that the five-point dilaton amplitude must vanish in six dimensions, which implies:
\eq
\Delta a=\frac{2b^2}{3f^4}>0\,.
\eqe
This gives a simple derivation of six-dimensional $a$-theorem obtained in~\cite{N20, US1, N10}.

For $\mathcal{N}=4$ supergravity in four dimensions, there is an anomaly for the U(1) subgroup of the global SU(1,1) symmetry, under which the two scalars of the theory transform as a doublet~\cite{N4Anom}. The scalars are the goldstone bosons of the coset SU(1,1)/U(1) and the presence of the anomaly reflects itself in the appearance of U(1) non-preserving amplitudes at one loop and beyond. The single-soft limits of the anomalous amplitudes were studied in~\cite{N4Radu}, and were shown to be non-vanishing and proportional to the U(1) generator acting on the lower point amplitude with one scalar removed. Here we will show that the behaviour is in fact universal, namely:
\eq
M_{\alpha,\beta,\pi}|_{q\rightarrow0}=\frac{1}{2}\sum_{i}G_{U(1)i}M_{\alpha,\beta}\,,
\eqe
regardless $M_{\alpha,\beta,\pi}$ is anomalous or not. Here $G_{U(1)i}$ is the single site U(1) symmetry generator, and $M_{\alpha,\beta}$ is the gravity scattering amplitude, stripped of its dimensionful coupling constant. This implies that the non-anomalous amplitudes will potentially have non-vanishing single soft scalar limits as well. We will use the five-point MHV one-loop amplitude to illustrate this property.

The above modified soft theorems can be derived using current correlators similar to that used to derive the original Adler's soft pion theorem~\cite{AdlerZero}. We will show that in the presence of anomalies, the single soft scalar limit is given as:
\eq
 M_{\alpha,\beta,\varphi}|_{q\rightarrow0}=\frac{i}{F}\left(-\langle \alpha| \delta \Gamma|\beta\rangle+q\cdot N_{\alpha,\beta}\right)\,.
\eqe
where $\delta \Gamma$ is the variation of the anomalous effective action, $\alpha,\beta$ are a set of initial and final states, $q$ is the momentum of the Goldstone boson $\varphi$ and $F$ the to be determined decay constant. On the other hand, the double soft limits are not modified. 

Finally we turn to one of the most important applications of scalar soft theorems: the compatibility of counter terms with duality symmetries in extended supergravity theories. Symmetries of operators are often notoriously difficult to verify as non-linear extension of symmetry transformations are necessary. On the other hand duality symmetries can be checked directly on the matrix elements generated by the operator by verifying whether or not they satisfy the requisite single and double soft theorems~\cite{Simple}. 

We will visit the case of three-loop candidate counter term $R^4$ in half-maximal supergravity, whose UV divergence was shown to be absent via explicit computation~\cite{N41}. We extract matrix elements of $R^4$ with half-maximal supersymmetry by considering the $\mathcal{O}(\alpha'^3)$ expansion of heterotic string amplitudes. Our computations have been greatly simplified by the realisation that the leading transcendental part of $\alpha'$ expansion of string theory amplitudes are in fact universal~\cite{alphaexpansion}. At $\alpha'^2$ this was already observed as the universality of operator $F^4$ in bosonic and superstring~\cite{Stieberger:2006te}. This allow us to directly extract matrix element of $R^4$, up to possible local polynomial ambiguity. This ambiguity will be removed by requiring the single soft-limit vanishes in accordance with preservation of U(1) duality symmetry. Once the ambiguity is removed, we can subject the completion to soft test at higher multiplicity. Importantly starts at seven points, it is possible to choose particular helicity configurations such that the results of single-soft limits are non-local, thus such choices would avoid the potential ambiguity of local terms. We will explicitly show that the completion fixed by the six-point soft constraint passes the soft tests at seven points.

At four loops, UV divergences are present~\cite{N42}, corresponding to $D^2R^4$ as well as anomalous operators $D^2R^3(D^2\phi)$ and $D^2R^2(D^2\phi)^2$. An interesting aspect of the result obtained in~\cite{N42} is that the coefficients of the divergences have universal transcendental structures. This result can be understood through soft theorems, as the soft limit of U(1) preserving matrix elements must be proportional to U(1) non-preserving ones, with the proportionality constant being an algebraic number.   

This paper is organised as follows. In section \ref{section:correlationfunction}, we begin with a brief introduction of soft theorems through correlation functions of conserved currents, and discuss the modifications necessary due to the presence of anomalies. In section~\ref{Dila}, we discuss modified soft theorems in the context of dilation effective action, as well as constraints implied by supersymmetry. In section \ref{N4Amp}, we discuss soft theorems for the anomalous U(1) duality symmetry of $\mathcal{N}=4$ supergravity with the focus on one-loop amplitudes. Finally, in section \ref{section:counterterm} we discuss the compatibility of $R^4$ with duality symmetry, as well as its implication for four-loop counter terms. We end with a summary of the results in section \ref{section:conclusion}.

%%%%%%%%%%%%%%%%%%%%%%%%%%%%%%%%%%%%%%%%%%%%%%%%%%%%%%%%%%%%%
\section{Soft limits for anomalous symmetries}  \label{section:correlationfunction}
%%%%%%%%%%%%%%%%%%%%%%%%%%%%%%%%%%%%%%%%%%%%%%%%%%%%%%%%%%%%%
%%%%%%%%%%%%%%%%%%%%%%%%%%%%%%%%%%%%%%%%%%%%%%%%%%%%%%%%%%%%%
\subsection{Single-soft limit}
%%%%%%%%%%%%%%%%%%%%%%%%%%%%%%%%%%%%%%%%%%%%%%%%%%%%%%%%%%%%%
In this section, we investigate soft theorems for theories related to symmetry breaking. We begin by reviewing Adler's soft theorem~\cite{AdlerZero}. Consider the matrix element where the current associated with  the broken generators are sandwiched between initial and final states $(\alpha,\beta)$:
\eq
\langle \alpha| J^\mu(x)|\beta\rangle\,.
\eqe
Since the current associated with the broken generator creates a Goldstone  boson from the vacuum, $\langle 0|J^{\mu}(0)|\pi(q)\rangle=iq^\mu F$, this matrix element must have a massless pole that reflects the creation of the goldstone boson. In other words 
\eq
\langle \alpha| J^\mu(0)|\beta\rangle=\frac{iq^\mu F}{q^2}M_{\alpha,\beta,\pi}+N^\mu_{\alpha,\beta}
\eqe
where $M_{\alpha,\beta,\pi}$ is the scattering amplitude for emitting a goldstone boson with $-q=p_\alpha+p_{\beta}$ during the transition $\alpha\rightarrow\beta$, while $N^\mu_{\alpha,\beta}$ is the pole free part of the matrix element. For spontaneous broken symmetries, the current is still conserved, and thus we have\footnote{The zero on the LHS can also be understood via Ward identity, and subsequent LSZ reduction.}  
\eq\label{Seq}
0=\langle \alpha| \partial\cdot J(x)|\beta\rangle=e^{iq\cdot x}q^\mu\langle \alpha| J^\mu(0)|\beta\rangle=e^{iq\cdot x}\left(i FM_{\alpha,\beta,\pi}+q\cdot N_{\alpha,\beta}\right)
\eqe
This leads to:
\eq
M_{\alpha,\beta,\pi}=\frac{i}{F}q\cdot N_{\alpha,\beta}
\eqe
Taking the momentum of the Goldstone boson to be near zero $q\rightarrow0$, we see that unless $N_{\alpha,\beta}$ also develops a singularity in this limit, the RHS of the above equation would vanish, indicating that the amplitude with one soft goldstone boson to vanish. The potential for $N_{\alpha,\beta}$ to develop a singularity is if the effective theory allows the Goldstone boson to couple via three-point interactions. In such cases, there will be diagrams where the Goldstone boson couples to the external leg:
$$\includegraphics[scale=0.5]{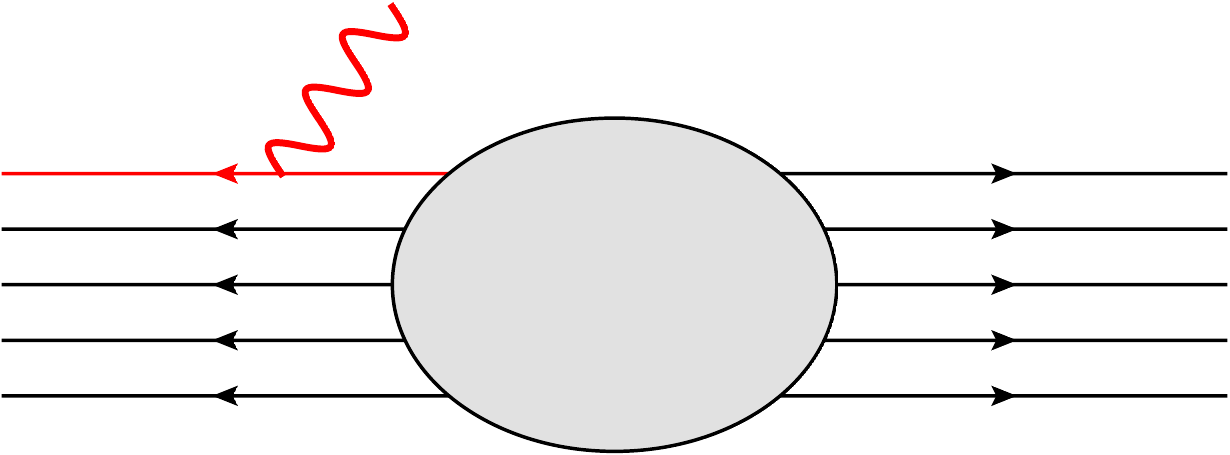}$$
For such diagrams, there will be a factor of $1/p\cdot q$ in $N_{\alpha,\beta}$ such that $q\cdot N_{\alpha,\beta}$ remains finite in the limit $q\rightarrow0$. For non-linear sigma models which are the low energy effective theory for spontaneous breaking of global symmetries, there are no three-point vertices and the above scenario would not arise. Thus one finds:
\eq
M_{\alpha,\beta,\pi}|_{q\rightarrow0}=0\,.
\eqe
If the Goldstone bosons couple to other massless degrees of freedom, as with the case of extended supergravity theories, the vanishing of the soft Goldstone boson limit can be understood from the fact that such soft limit probes the vacuum moduli space~\cite{Simple}.

We now consider the case when the symmetry is broken via an anomaly. The presence of an anomaly implies that the classical action is modified by an effective action that varies into the anomaly:\
\eq
\delta \Gamma=Anom\,.
\eqe
This modifies eq.(\ref{Seq}) as follows:
\eq
\langle \alpha| \partial\cdot J(x)|\beta\rangle=\langle \alpha| \delta \Gamma|\beta\rangle=e^{iq\cdot x}\left(i FM_{\alpha,\beta,\pi}+q\cdot N_{\alpha,\beta}\right)\,.
\eqe
Thus the original soft theorem is modified by a non-vanishing matrix element that is generated by the variation of the effective action. Note that if the initial and final states $\alpha,\beta$ are such that $\langle \alpha| \delta \Gamma|\beta\rangle=0$, then the soft theorem is not modified. An example where $\langle \alpha| \delta \Gamma|\beta\rangle\neq0$ will be the dilaton effective action, as we will show in the next section. In summary, for anomalous symmetries, the soft theorem is given via
\eq\label{Result}
 M_{\alpha,\beta,\pi}|_{q\rightarrow0}=\frac{i}{F}\left(-\langle \alpha| \delta \Gamma|\beta\rangle+q\cdot N_{\alpha,\beta}\right)\,.
\eqe
where $\Gamma$ is the anomalous effective action.

In section \ref{Dila} and \ref{N4Amp}, we will discuss implications of eq.(\ref{Result}) in two setups, the dilation effective action and the anomalous U(1) duality symmetry of $\mathcal{N}=4$ supergravity.

%%%%%%%%%%%%%%%%%%%%%%%%%%%%%%%%%%%%%%%%%%%%%%%%%%%%%%%%%%%%%
\subsection{Double-soft limit} \label{section:double-soft}
%%%%%%%%%%%%%%%%%%%%%%%%%%%%%%%%%%%%%%%%%%%%%%%%%%%%%%%%%%%%%
In this subsection we discuss general double soft limits for scalars and fermions. The double-soft limit explores the group theory structure of spontaneous broken symmetry. A more detailed discussion of double soft limits can be found in the appendix of~\cite{Jaro}. Consider the coset $G/H$ with generators $X, T$, where $X$'s are the broken generators. They satisfy the following algebra: 
\eq
[T^a,T^b]=f^{ab}\,_cT^c,\quad [T^a,X^I]=f^{aI}\,_JX^J,\quad [X^I,X^J]=f^{IJ}\,_a T^a\,.
\eqe 
The associated current for $X^I$ and $T^a$ will be denoted as $J^I_\mu$ and $J^a_{\mu}$ respectively. We will consider the following matrix element:
\eq\label{Initial}
q_1^\mu q_2^\nu\langle J_{\mu}^I(q_1)J_\nu^J(q_2) J_{\rho_1}^{K_1}(p_1)\cdots J_{\rho_n}^{K_n}(p_n)\rangle\,.
\eqe
Apply the Ward identity and perform an LSZ reduction on legs $1,\cdots,n$, one obtains 
\eqa\label{MidTerm}
\nonumber&& q_1^\mu q_2^\nu\langle J_{\mu}^I(q_1)J_\nu^J(q_2) J_{\rho_1}^{K_1}(p_1)\cdots J_{\rho_n}^{K_n}(p_n)\rangle\bigg|_{\rm LSZ}\\
&=&-\frac{1}{2}f^{IJ}\,_a(q_1-q_2)^{\mu}\langle J_{\mu}^a(q_1+q_2)|\pi^{K_1}(p_1)\cdots \pi^{K_n}(p_n)\rangle\left(\prod_{j=1}^niF\,p_{j\rho_j}\right)\,.
\eqae
As we will take momenta $q_1,q_2$ to be soft, the RHS will be non-vanishing only if there is a singular contribution from $\langle J_{\mu}^a(q_1+q_2)|\pi^{K_1}(p_1)\cdots \pi^{K_n}(p_n)\rangle$. Just as the single soft discussion, the most singular contribution stems form attaching $J_{\mu}^a$ to an external scalar line. This contribution can be deduced from the Ward identity:
\eq
q^\mu\langle J_{\mu}^a(q)\pi^{K_1}(p_1)\cdots \pi^{K_n}(p_n)\rangle=\sum_{i=1}^nf^{K_iJa}\langle \pi^{K_1}(p_1)\cdots \pi^{J}(p_i)\cdots \pi^{K_n}(p_n)\rangle\,.
\eqe
At leading order in $q$ the solution is:
\eq
\langle J_{\mu}^a(q)\pi^{{I_1}}(p_1)\cdots \pi^{I_n}(p_n)\rangle{=}\sum_{i=1}^n\frac{f^{I_iJa}p_{i\mu}}{q\cdot p_i}\langle \pi^{I_1}(p_1)\cdots \pi^{J}(p_i)\cdots \pi^{I_n}(p_n)\rangle{+}\mathcal{O}(q^0)
\eqe
Substituting into the RHS of eq.(\ref{MidTerm}) one finds the following finite contribution:
\eq
-f^{IJ}\,_a\sum_{i=1}^n\frac{f^{K_iJa}(p_i\cdot(q_1-q_2))}{2(p_i\cdot(q_1+q_2))}\langle \pi^{K_1}(p_1)\cdots \pi^{J}(p_i)\cdots \pi^{K_n}(p_n)\rangle\left(\prod_{j{=}1}^niF\,p_{j\rho_j}\right)\,.
\eqe
Finally direct LSZ reduction on of the LHS of eq.(\ref{Initial}) and taking the double soft-limit one finally finds:
\eqa
\nonumber && \langle \pi^{I}(q_1)\pi^{J}(q_2) \pi^{K_1}(p_1)\cdots \pi^{J}(p_i)\cdots \pi^{K_n}(p_n)\rangle\bigg|_{q_1,q_2\rightarrow0}\\
&=&-f^{IJ}\,_a\sum_{i=1}^n\frac{f^{K_iJa}(p_i\cdot(q_1{-}q_2))}{2(p_i\cdot(q_1{+}q_2))}\langle \pi^{K_1}(p_1)\cdots \pi^{J}(p_i)\cdots \pi^{K_n}(p_n)\rangle\,.
\eqae
Now in the presence of anomalies, one would have an extra term on the RHS of eq.(\ref{MidTerm}) since the Ward identity is modified. However, this extra term does not yield the necessary poles to leave finite contributions after LSZ reduction, and hence does not contribute. \textit{Thus if the spontaneously broken symmetry is in fact anomalous, while the single soft limit is modified, the double soft theorem is not. } For example, stringy corrections to maximal supergravity breaks the isotropy group SU(8). As a consequence, the single soft limit no longer vanishes~\cite{ElvangSUSYCT}, although the double soft-limit is exactly the same as $\mathcal{N}=8$ supergravity, as we will demonstrate.

The above result can be straightforwardly extended to fermionic symmetries. For example, for Goldstinos associated with spontaneously broken supersymmetry, the current is now $J_{\mu}^{\alpha}, J_{\nu}^{\dot{\alpha}}$. The Ward identity generated from two currents now yields:\footnote{Double soft limits of Goldstinos were studied in~\cite{deWit}. Here we present a more modern presentation.}
\eq\label{FermiInitial}
q_1^\mu q_2^\nu\langle J_{\mu}^\alpha (q_1)\tilde{J}_\nu^{\dot{\alpha}}(q_2) \cdots \rangle=\frac{(q_1-q_2)^{\mu}}{2}\langle 1|\gamma^\nu|2]\langle \theta_{\mu\nu}\,^{\alpha\dot{\alpha}}(q_1+q_2)\cdots\rangle\,,
\eqe
where $\theta_{\mu \nu}\,^{\alpha\dot{\alpha}}$ is the stress tensor, and $\cdots$ are the remaining Goldstinos. The presence of the pre-factor $\langle 1|\gamma^\nu|2]$ is due to the fact that the super current $J_{\mu}^{\alpha}$ excites a Goldstino from the vacuum,
\eq
\langle 0|J_{\mu}^{\alpha}|\chi\rangle =iF\gamma^\mu \lambda^\alpha
\eqe
where $\chi$ is an on-shell Goldstino state, $F$ the Goldstino decay constant, and $\lambda$ the external line factor. This then implies that the RHS of the above equation must carry the little group weights of the two Goldstinos. Again as we take $q_1,q_2\rightarrow 0$, only the singular terms in  $\langle \theta_{\mu\nu}(q_1+q_2)\cdots\rangle$ survive. The singular term is derived from the following Ward identity: 
\eq
q^\mu\langle \theta_{\mu\nu}(q)\cdots\rangle=\sum_{i=1}^np_{i\nu}\langle \cdots\rangle\rightarrow \langle \theta_{\mu\nu}(q)\cdots\rangle=\sum_{i=1}^n\frac{p_{i\nu}p_{i\mu}}{p_i\cdot q}\langle \cdots\rangle+\mathcal{O}(q^0)\,.
\eqe
Putting the above into eq.(\ref{FermiInitial}) and performing LSZ reduction one finds
\eqa
\nonumber\langle\chi(q_1)\bar{\chi}(q_2)\chi(p_1)\cdots\chi(p_n)\rangle|_{q_1,q_2\rightarrow0}{=}\sum_{i=1}^n\frac{p_{i}\cdot(q_1{-}q_2)}{2p_i\cdot (q_1{+}q_2)}\langle 1|p_i|2]\langle \chi(p_1)\cdots\chi(p_n)\rangle\,,
\eqae
This is precisely what was found in~\cite{US3}.

In~\cite{US3}, it was noted that the double soft-limit of fermions in supergravity theories also admit universal double soft-fermion limits, although there is no known broken symmetries associated with the massless spin$-\frac{1}{2}$ particles. In particular one finds:\footnote{Here, only the four-dimensional version is presented. There is also a similar soft theorem for three-dimensional supergravity, which is applicable to all fermions in the multiplet.}
\eq \label{eq:doublefermion}
\langle \psi^1(q_1) \bar{\psi}^2 (q_2)\psi(p_1)\cdots\psi(p_n)\rangle|_{q_1,q_2\rightarrow0}{=}\sum_{i=1}^n\frac{\langle 1|p_i|2]}{2p_i\cdot (q_1{+}q_2)}R_i\langle \psi(p_1)\cdots\psi(p_n)\rangle
\eqe
where two soft fermions $\psi^1$ and $\bar{\psi}^2$ are spin$-\frac{1}{2}$ fermions with one and only one overlap R-symmetry index, and $R_i$ is the isotropy group of the coset space which the scalars of the theory parametrizes. Compared with the previous derivation of Goldstino soft-limits, the above soft theorem appears to imply the existence of spinor operators, $\mathcal{O}^{\alpha}, \tilde{\mathcal{O}}^{\dot{\alpha}}$ such that the anti-commutator generates 
\eq
\{\mathcal{O}^{\alpha},\tilde{\mathcal{O}}^{\dot{\alpha}}\}\sim(J_R)^{\alpha\dot{\alpha}}
\eqe
where $(J_R)^{\alpha\dot{\alpha}}=\sigma_\mu^{\alpha\dot{\alpha}}J_R^\mu$ is the isotropy group current. Further more there is an equivalence between 
\eq
\langle\psi^1(q_1) \bar{\psi}^2(q_2)\psi(p_1)\cdots\psi(p_n)\rangle|_{q_1,q_2\rightarrow0}{=} \langle  \mathcal{O}^{\alpha}(q_1)\tilde{\mathcal{O}}^{\dot{\alpha}} (q_2) \psi(p_1)\cdots\psi(p_n)\rangle|_{q_1,q_2\rightarrow0}.
\eqe 
While so far we do not have a symmetry principle that underlines the above relations, we will proceed and test whether these soft fermion relations are also respected for string amplitudes where R-symmetry is broken by $\alpha'$ corrections.
%%%%%%%%%%%%%%%%%%%%%%%%%%%%%%%%%%%%%%%%%%%%%%%%%%%%%%%%%%%%%
\section{Amplitudes of dilation effective action}\label{Dila}
%%%%%%%%%%%%%%%%%%%%%%%%%%%%%%%%%%%%%%%%%%%%%%%%%%%%%%%%%%%%%
The dilaton effective action is constructed by considering the most general diffeomorphism invariant action, that yields the trace anomaly upon Wely transformation, where the dilaton shifts as $\tau\rightarrow \tau+\sigma$. The trace anomaly generally takes the form
\eq
\langle T^\mu\,_\mu\rangle=\sum_{i}c_iI_i-(-)^{\frac{d}{2}}aE_d
\eqe
where $E_d$ is the even $d$-dimensional Euler density and $\sqrt{-g}I_i$ are conformal invariants. It was proposed that the coefficient $a$ serves as a candidate function that monotonously decreases along a RG flow from UV to IR~\cite{Cardy}. In the IR, the massless degrees of freedom includes a dilaton $\tau$ which is the Goldstone boson if conformal symmetry is spontaneously broken,\footnote{There are $4+1$ broken generators when conformal symmetry is spontaneously broken, however there is only one Goldstone boson. The mismatch is due to the broken symmetry is a space-time symmetry. See~\cite{Low} for a detailed discussion.} or it is a compensating field if conformal symmetry is explicitly broken. The dynamics of the dilaton is then governed by the dilaton effective action, which is defined as a diffeomorphism invariant functional $\Gamma[g,\tau]$, such that under a Weyl transformation one generates the anomaly:
\eq
\delta \Gamma[g,\tau]_{dila}=\int d^d x\sqrt{-g}\left(\sum_{i}c_iI_i-(-)^{\frac{d}{2}} \Delta a E_d\right)\,.
\eqe
From the above one sees that the $\Delta a$ appears in the coefficient of the $d$-derivative terms in dilaton effective action in the IR, i.e. it appears in the coefficient of the mass-dimension $d$ amplitudes. The explicit form of the dilaton effective action is given in~\cite{Elvang:2012yc}. We now study whether soft theorems we proposed are satisfied, and impose any non-trivial constraints on $\Delta a$?

The three-point amplitude for dilaton scattering vanishes from Lorentz invariance. This implies that there are no singular terms for $ N_{\alpha,\beta}$ in eq.(\ref{Result}) and thus  $q\cdot N_{\alpha,\beta}=0$ as $q\rightarrow0$. Hence naively one would expect that for the dilaton effective action,  
\eq\label{Result2}
M_{\alpha,\beta,\pi}|_{q\rightarrow0}=\langle \alpha| \delta \Gamma|\beta\rangle=\sum_{i}\mathfrak{D}_iM^{\Gamma}_{\alpha,\beta}\,,
\eqe
where $\mathfrak{D}_i$ is the single site dilatation operator, and $M^{\Gamma}_{\alpha,\beta}$ is the matrix element generated from the anomalous effective action.  However $\tau$ is not the physical dilaton. Rather, it is related to the physical dilaton $\varphi$ via:
\eq
e^{-\frac{d-2}{2}\tau}=1-\frac{\varphi}{f^{(d-2)/2}}\quad\rightarrow\quad\tau= \frac{2}{d-2}\frac{\varphi}{f^{(d-2)/2}}\,,
\eqe 
where the RHS holds at linear order, and $f$ is the mass scale. Thus shifting the physical dilaton field $\varphi$ by $\sigma'$ corresponds to shifting the dilaton field $\tau$ by $\sigma=\frac{2}{d-2}\frac{\sigma'}{f^{(d-2)/2}}$. Adjusting for the additional proportionality constant, we find that the soft-theorem for dilaton effective action is given as:
\eq\label{DilaResult}
 M_{\alpha,\beta,\varphi}|_{q\rightarrow0}=\frac{2}{d-2}\frac{1}{f^{(d-2)/2}}\sum_{i}\mathfrak{D}_iM^{\Gamma}_{\alpha,\beta}\,.
\eqe
Note that with the extra factors of $1/f^{(d-2)/2}$ in eq.(\ref{DilaResult}) indicates that the soft behavior of the $n$-point $k$-derivative amplitude is proportional to the $(n{-}1)$-point $k$-derivative amplitude. Let us now verify the above dilaton soft theorem in explicit examples. 

%%%%%%%%%%%%%%%%%%%%%%%%%%%%%%%%%%%%%%%%%%%%%%%%%%%%%%%%%%%%%
\subsection{Soft limits for dilaton effective action}
%%%%%%%%%%%%%%%%%%%%%%%%%%%%%%%%%%%%%%%%%%%%%%%%%%%%%%%%%%%%%
First let's consider the dilaton effective action derived from the gravity dual of a single D-brane separated from a stack of $N$  D-branes. This is dual to the Coloumb branch of a CFT. In the large $N$ limit, this is described by that of a probe brane in the supergravity background, whose dynamics is governed by the $d$-dimensional DBI action, 
\eq
S_{\rm DBI}=-\int d^{d}x\frac{1}{|z|^d}\sqrt{1+(\partial z)^2}
\eqe 
where the scalars $\phi^i$ form the vector $\vec{z}$, one of which is the dilaton. To compute amplitudes, one needs to give one of the scalars a vev and expand  $|z|=f^{(d-2)/2}+\cdots$. The leading term in the expansion will simply be the flat space DBI, which has only even multiplicity S-matrix. Note that according to eq.(\ref{DilaResult}), the vanishing of odd-multiplicity S-matrix element implies that the single soft-limit vanishes, which is indeed the case~\cite{Jaro1}.\footnote{In fact, it vanishes faster than the usual Adler's zero.}

We can also directly verify the above result for the dilaton amplitude in 6D, up to $\mathcal{O}(p^6)$ they are given as~\cite{Elvang6D}  
\eqa\label{AmpList}
\nonumber M_4^{p^4}=\frac{b}{2f^8}(s^2+t^2+u^2),&&\quad M_4^{p^6}=\left[\frac{3\Delta a}{2}-\frac{b^2}{f^4}\right]\frac{3}{f^8}stu\\
M^{p^4}_5=\frac{3b}{4f^{10}}\sum_{1\leq i<j\leq5}s^2_{ij},&&\quad M_5^{p^6}=\left[\frac{3\Delta a}{2}-\frac{b^2}{f^4}\right]\frac{2}{f^{10}}\sum_{1\leq i<j\leq5}s^3_{ij}
\eqae
where the superscript on each amplitude indicates the degree of derivative coupling for the associated interacting vertex, and the coefficient $b$ is a parameterization of the freedom to add Weyl invariant terms to the action~\cite{Elvang6D}. We have also listed and computed the six and seven-point amplitudes in the appendix~\ref{App}. The single soft-limit of these amplitudes satisfies the following universal behavior:
\eqa
\nonumber \quad M^{p^4}_n\rightarrow { n{-}2 \over f^2 }M^{p^4}_{n{-}1} \\
\quad M^{p^6}_n\rightarrow { n{-}1 \over f^2 } M^{p^6}_{n{-}1}\,.
\eqae
The factor of $f^{-2}$ gives the correct dimension compensation factor for $d=6$, as can be seen from eq.(\ref{DilaResult}). Note that this result holds for all values of $\Delta a $ and $b$, and thus holds for both explicit and spontaneous symmetry breaking. We can also test results in other dimensions. The dilaton effective action in $d=2k$ dimensions was derived in~\cite{Elvang:2012yc}, and explicitly amplitudes were listed up to 8-points for four, six and eight derivatives. We've performed an exhaustive check and find that the universal soft theorem for dilaton effective action is given by:
\eq\label{SoftFactor}
M_n\rightarrow { S_n \over f^{(d-2)/2} }M_{n{-}1},\quad S_n=n-3+\frac{2\Delta-4}{d-2} 
\eqe
where $\Delta$ is the number of derivatives of the amplitude. To see that eq.(\ref{SoftFactor}) is equivalent to eq.(\ref{DilaResult}), note that in $d$ dimensions, the on-shell dilatation operator is given by 
\eq\label{DDef}
\mathfrak{D}=\sum_{i=1}^n \left(m_i\frac{\partial }{\partial m_i}+\frac{d-2}{2}\right)
\eqe 
where $\sum_i m_i\frac{\partial }{\partial m_i}$ is an operator that counts the mass dimension of the amplitude. In terms of spinor helicity it is given as:
\eq
D=4:\; \frac{1}{2}\left(\lambda_i^\alpha \frac{\partial}{\partial \lambda_i^\alpha}+\tilde\lambda_i^{\dot{\alpha}} \frac{\partial}{\partial \tilde\lambda_i^{\dot\alpha}}\right),\quad D=6:\; \frac{1}{2}\left(\lambda_i^{Aa} \frac{\partial}{\partial \lambda_i^{A a}}+\tilde\lambda_{iA\dot{a}} \frac{\partial}{\partial \tilde\lambda_{iA\dot{a}}}\right) \, .
\eqe 
Now the mass-dimenion of an $\Delta$ derivative amplitude is $\Delta -d$, where the extra $-d$ comes from the momentum conservation delta function $\delta^d(P)$. Thus the soft factor in eq.(\ref{SoftFactor}) can be rewritten as 
\eq
S_n=\frac{2}{d-2}\left[(n-3)\frac{d-2}{2}+\left(\sum_{i=1}^{n-1} m_i\frac{\partial }{\partial m_i}\right)+d-2\right]=\frac{2}{d-2}\mathfrak{D}
\eqe 
where $\mathfrak{D}$ is given in eq.(\ref{DDef}) with the summation over $1,\cdots, n{-}1$. This confirms that in general the soft-limit of the dilaton effective action, is given by 
\eq
M_n\rightarrow \frac{2}{d-2}{1 \over f^{(d-2)/2}}\mathfrak{D} M_{n{-}1}\,,
\eqe
in accordance with eq.(\ref{DilaResult}).
 
%%%%%%%%%%%%%%%%%%%%%%%%%%%%%%%%%%%%%%%%%%%%%%%%%%%%%%%%%%%%%
\subsection{Supersymmetry Constraints}
%%%%%%%%%%%%%%%%%%%%%%%%%%%%%%%%%%%%%%%%%%%%%%%%%%%%%%%%%%%%%
As illustrated above, soft theorems are satisfied by the dilaton effective action and impose no constraints on either $\Delta a$ or $b$. Constraints on $\Delta a$ in four dimensions were achieved by combining dispersion relations with the optical theorem~\cite{KS}. In six dimensions, constraints can be found instead by considering supersymmetry~\cite{N20, US1, N10}. For completeness, we will derive the six-dimensional SUSY constraints in a much simpler fashion. (This subsection is somewhat slightly outside the mainline of this paper.)

In six dimensions, supersymmetry is categorised by the number of chiral and anti-chiral supercharges, $(\tilde{n}_L,\tilde{n}_R)$, each carrying 4 complex components. The superconformal algebra is $OSp(8|2\tilde{n})$, indicating the corresponding supersymmetry is chiral. We will suppress the subscript $L$ from now on. In terms of on-shell variables the $4n$ supersymmetry charges are given by:
\eq
Q^{AI}=\sum_{i=1}^n\lambda_i^{Aa}\eta^{I}_{ia},\quad \tilde{Q}^{A}_{I}=\sum_{i=1}^n\lambda_i^{Aa}\frac{\partial}{\partial \eta^I_{ia}}
\eqe
where $A\in$ SU(4) Lorentz symmetry and $a\in$ SU(2) little-group and $I=1,\cdots,\tilde{n}$.\footnote{The chiral half of six-dimensional SO(4)$\sim$SU(2)$\times$SU(2) little group } The bosonic and fermionic variables carry the kinematic and multiplet degrees of freedom respectively~\cite{Donal, Warren}. For example the momentum $P_i^{AB}$ is given by $\lambda^{Aa}_{i}\lambda^{B}_{ia}$. 

Except the maximal supersymmetry, the on-shell degrees of freedom are contained in the two separate on-shell multiplets:
\eqa\label{MultipletLs}
\nonumber\mathcal{N}=(2,0): ~ \quad \quad \quad \quad \Phi(\eta^I)&=&\phi+\cdots+\eta_{(a}^{I}\eta_{b)I}B^{ab}+\cdots (\eta^1)^2(\eta^2)^2\bar{\phi}\\
\nonumber\mathcal{N}=(1,0):  \quad {\rm tensor} ~~ \Psi^a(\eta)&=&\psi^a+\eta^a\phi+\eta^bB_b\,^a+(\eta)^2\bar{\psi}^a,\quad a=1,2\\
 \nonumber \;{\rm hyper} ~~~~ \Phi(\eta)&=&\phi+\eta^a\psi_a+(\eta)^2\phi'\\
  \nonumber  \bar\Phi(\eta)&=&\bar\phi+\eta^a\bar\psi_a+(\eta)^2\bar\phi'\\
   \;{\rm vector} ~~ \Psi^{\dot{a}}(\eta)&=&\psi^{\dot{a}}+\eta^{b}A_a\,^{\dot{a}}+(\eta)^2\bar{\psi}^{\dot{a}}
\eqae
where $B^{ab}$ is the self-dual two-form and $A^{a\dot{a}}$ is a one-form.

The most relevant generator in our discussion is the R-symmetry generators. For $\mathcal{N}=(\tilde{n},0)$ supersymmetry, the Sp(2$\tilde{n}$) R-symmetry is represented non-linearly on-shell as~\cite{6DTwistor},
\eq
R^{I}\,_{J}=\eta^{I}\cdot\frac{\partial}{\eta^J}-\delta^I\,_J,\quad R^{(IJ)}=\eta^{I}\cdot \eta^{J},\quad R_{(IJ)}=\frac{\partial}{\partial \eta^{I}}\cdot \frac{\partial}{\partial \eta^{J}}\,.
\eqe
Here we use $\cdot$ to represent summation over both the index $i$ and little-group index $a$. It is due to the latter that $R^{(IJ)}$ and $R_{(IJ)}$ is symmetric. Note that due to the constant $-1$ in the generators $R^{1}_1$ and $R^{2}_2$, it requires that the amplitude must be of degree $n$ in $\eta^{I}$. Combined with the multiplicative susy constraint, we find that in general the supersymmetric amplitude can be written as 
\eq
\delta^{4\tilde{n}}(Q)f(\eta^{I},\lambda)
\eqe 
where $f$ is an homogenous polynomial of $\eta^J$ with degree $n-4$. The function $f$ must also carry little group indices if the external legs are part of the tensor or vector multiplet as listed in eq.(\ref{MultipletLs}).

In the following, we will consider the possible supersymmetric completions for the six-derivative five-point amplitude. We will only consider local completions, since there are no $\mathcal{N}=(1,0)$ respecting residues for a factorization pole. To see this, note that for a pole to be present at four or five-point, the residue must be the direct product of a three-point and a four-point amplitude, with four derivatives each. For $\mathcal{N}=(1,0)$ the three-point amplitude is given by
\eq
\mathcal{M}_3\sim [(u^a_1\eta_{1a})(u^a_2\eta_{2a})(w^a_3\eta_{3a})+{\rm cyclic}]\times f(u_{ia},\tilde{u}_{i\dot{a}})
\eqe
where $u,\tilde{u},w$ are special variables for degenerate three-point kinematics~\cite{Donal}.\footnote{They are defined through:
$$\lambda^A_{ia}\tilde{\lambda}_{Ai{+}1\dot{a}}\equiv u_{ia}\tilde{u}_{i{+}1\dot{a}}$$
and 
$$u_{i[a}w_{ib]}\equiv \epsilon_{ab}\,.$$} As $f$ must have mass dimension $7/2$, this is only possible if the gravitational multiplet is included, and thus ruled out in this discussion.

%%%%%%%%%%%%%%%%%%%%%%%%%%%%%%%%%%%%%%%%%%%%%%%%%%%%%%%%%%%%%
\subsubsection{$\mathcal{N}=(2,0)$}
%%%%%%%%%%%%%%%%%%%%%%%%%%%%%%%%%%%%%%%%%%%%%%%%%%%%%%%%%%%%%
First, consider $\mathcal{N}=(2,0)$. A Local ansatz that satisfies the multiplicative $\mathcal{N}=(2,0)$ for $\mathcal{A}_5$ is 
\eq
\mathcal{N}=(2,0):\quad \mathcal{M}_5=\delta^8(Q)(\alpha \langle q^I_1 p_2 q^J_1\rangle+\beta \langle q^I_1 p_2 q^J_3\rangle)+ {\rm perm} (1,2,3,4,5)\, ,
\eqe
where $I\neq J$. The extra $q^I$ factors are required since R-symmetry enforces the function to be degree $5$ in $\eta^1$ and $\eta^2$. Note that after permutations, the ansatz vanishes due to momentum conservation. Thus one concludes that for $\mathcal{N}=(2,0)$, 
\eq
\Delta a=\frac{2b^2}{3f^4}.
\eqe
In agreement with~\cite{Elvang6D}. 
%%%%%%%%%%%%%%%%%%%%%%%%%%%%%%%%%%%%%%%%%%%%%%%%%%%%%%%%%%%%%
\subsubsection{$\mathcal{N}=(1,0)$}
%%%%%%%%%%%%%%%%%%%%%%%%%%%%%%%%%%%%%%%%%%%%%%%%%%%%%%%%%%%%%
Now for $\mathcal{N}=(1,0)$, let us first consider the case where the dilaton lies in the tensor multiplet. Since the entire multiplet is fermionic, as indicated by the leading component field in eq.(\ref{MultipletLs}), the corresponding ansatz must be totally anti-symmetric and must carry a little group index on each leg. The ansatz that satisfies these conditions would be 
\eqa
\mathcal{N}=(1,0):\quad 
\mathcal{M}_5 &=&
\delta^4(Q) \big( \alpha  \langle q_1p_2 3_{c}\rangle\langle1_a2_b 4_d5_e\rangle 
+
\beta  \langle q_1 2_b 3_{c} 4_d \rangle\langle 1_a p_2 5_e\rangle \big) \cr
&+& (-){\rm perm} (1,2,3,4,5)
\eqae
where $(-)$ indicates that the permutation is multiplied by the signature of the permutation. To respect the full $\mathcal{N}=(1,0)$, that the above ansatz has to satisfy $\tilde{Q}^A\mathcal{M}_5=0$. It is straightforward to show that 
\eqa
\tilde{Q}^A\mathcal{M}_5
&=&
\delta^4(Q) \lambda_1^{Af} 
\big( \alpha  \langle1_f p_2 3_{c}\rangle\langle1_a2_b 4_d5_e\rangle
+
\beta  \langle 1_f 2_b 3_{c} 4_d \rangle\langle 1_a p_2 5_e\rangle \big) \cr
&+& (-){\rm perm} (1,2,3,4,5) 
\neq0\,.
\eqae
Hence one again concludes that the amplitude cannot be supersymmetrized, and we have:
\eq
\Delta a=\frac{2b^2}{3f^4}.
\eqe
Finally, in passing we mention that for the case where the dilaton is in the vector multiplet, one arrives at the same result. We discuss this in detail in appendix~\ref{Vector}

%%%%%%%%%%%%%%%%%%%%%%%%%%%%%%%%%%%%%%%%%%%%%%%%%%%%%%%%%%%%%
\section{Soft constraints on anomalous $\mathcal{N}=4$ supergravity}\label{N4Amp}
%%%%%%%%%%%%%%%%%%%%%%%%%%%%%%%%%%%%%%%%%%%%%%%%%%%%%%%%%%%%%
For extended supergravity theories with $\mathcal{N}\geq4$, besides the usual R-symmetry, the equations of motions respects additional non-compact global symmetries. The massless scalars in the spectrum can be identified as parameterizing the coset manifold that arrises from spontaneously breaking the non-compact global symmetry. The remaining invariant subgroup exchanges the self-dual and anti-self dual abelian field strengths, and thus acquiring the name duality symmetry.  Such symmetries can be broken either by quantum corrections or by the presence of higher dimensional operators such as that implied by the string theory completion.  For the former, $\mathcal{N}=4$ supergravity is known to have an anomalous abelian subgroup of the SU(1,1) duality group~\cite{N4Anom}. Due to this anomaly, U(1) non-conserving amplitudes are present at one loop as  shown in~\cite{N4Radu}. In this section, we will study the fate of soft limits in the presence of the U(1) anomaly.

The on-shell degrees of freedom for $\mathcal{N}=4$ supergravity are captured in two multiplets:
\eqa\label{Super}
\nonumber \Phi&=&h^{+2}+\eta^I\psi^{+\frac{3}{2}}_I+\frac{\eta^I\eta^J}{2}A^{+1}_{IJ}+\frac{\eta^{3}_I}{3!}\lambda^{+\frac{1}{2}I}+\frac{\eta^4}{4!}\phi\\
\bar{\Phi}&=&\bar{\phi}+\eta^I\bar{\lambda}^{-\frac{1}{2}}_I+\frac{\eta^2_{KL}}{2}\bar{A}^{-1KL}+\frac{\eta^{3}_I}{3!}\bar{\psi}^{-\frac{3}{2}I}+\frac{\eta^4}{4!}h^{-2}
\eqae
The states can be obtained by tensoring pure Yang-Mills and $\mathcal{N}=4$ super Yang-Mills. The tree-level amplitude of this theory can be obtained though KLT relations~\cite{KLT}, or equivalently the BCJ double copy formalism~\cite{BCJ}, where the latter is applicable at loop level. This implies that in practice, the choice of $(\Phi,\bar{\Phi})$multiplet for each external leg is determined by the its helicity on the YM copy, i.e. a leg that has a plus (negative) helicity gluon correspond do the $\Phi$ ($\bar{\Phi}$) multiplet after tensoring. 
\eq
\langle+,+,\cdots,+\rangle\otimes {\rm SYM} =\langle \Phi,\Phi,\cdots,\Phi\rangle,\quad \langle-,+,\cdots,+\rangle\otimes {\rm SYM} =\langle \bar{\Phi},\Phi,\cdots,\Phi\rangle\,.
\eqe  
Following the notation of~\cite{N4Radu} the amplitudes of $\mathcal{N}=4$ supergravity can be denoted as N$^{\rm k}$MHV$^{(p,q)}$ where $p,q$ is the number of $\Phi$ and $\bar{\Phi}$ muiitplet respectively and N$^{\rm k}$MHV indicates the helicity structure for the $\mathcal{N}=4$ SYM copy.

 %%%%%%%%%%%%%%%%%%%%%%%%%%%%%%%%%%%%%%%%%%%%%%%%%%%%%%%%%%%%%
\subsection{U(1) duality symmetry and anomalous  amplitudes}
%%%%%%%%%%%%%%%%%%%%%%%%%%%%%%%%%%%%%%%%%%%%%%%%%%%%%%%%%%%%%

The two scalars of the theory can be thought of as parameterizing the coset SU(1,1)/U(1), where all fields are potentially charged under the U(1) isotropy group. The identification of the states via double copy is useful in determining their charges:\
\eq
q_{\rm U(1)}=(h({\rm YM})-h({\rm SYM}))\,
\eqe
where $h({\rm YM})$ and $h({\rm SYM})$ indicate the helicities of the states in YM and SYM copy respectively. One then finds the charge of each field is given as
 \eq
 \Psi^{\pm\frac{3}{2}}= \left(\pm \frac{1}{2}\right),\;\; A^{\pm1} = \left(\pm 1\right),\;\;\lambda^{\pm\frac{1}{2}}=\left(\pm \frac{3}{2}\right),\;\;\phi=(2),\;\;\bar{\phi}=(-2) \, .
 \eqe
Since under a U(1) rotation, one interchanges the positive and negative helicity photons, the symmetry is often referred to as duality symmetry that rotates the electric-magnetic field strengths. The constraint of vanishing total U(1) charge requires that both the YM and SYM amplitudes have the same helicity structure. Indeed tensoring tree-level amplitudes of distinct helicity vanishes in the KLT construction. Only N$^{\rm k}$MHV$^{(n{-}k{-}2, k{+}2)}$ are non-vanishing at tree level.

At loop level due to the anomaly, the above constraint no longer applies.  Explicit computation has shown found~\cite{N4Radu}
\eqa\label{N4Amps}
\nonumber \overline{\rm MHV}^{(n,0)}:\quad \mathcal{M}_n&=&\frac{i}{4\pi^2}(n{-}3)!\left(\frac{\kappa}{2}\right)^{n}\delta^8(\tilde{Q})\\
\nonumber {\rm MHV}^{(3,1)}:\quad \mathcal{M}_4&=&\frac{i}{4\pi^2}\left(\frac{\kappa}{2}\right)^{4}\frac{[23][34][42]}{\langle 23\rangle\langle 34\rangle\langle42\rangle}\delta^8(Q)\\
{\rm MHV}^{(5,0)}:\quad \mathcal{M}_4&=&\frac{i}{4\pi^2}\left(\frac{\kappa}{2}\right)^{5}\left(\sum_{S_5}\frac{\gamma^2_{ij}}{12 s_{ij}}\right)\delta^8(Q)
\eqae
 where $\tilde{Q}=\sum_i\tilde{\lambda}\bar{\eta}$, is the conjugate supercharge and 
 \eq
 \gamma_{12}=\frac{[12]^2[34][45][53]}{\langle12\rangle[23]\langle35\rangle[51]-[12]\langle23\rangle[35]\langle51\rangle}\,.
 \eqe
From eq.(\ref{Super}) we identify that $\eta^{I}$ carries $-\frac{1}{2}$ U(1) charge. The super amplitude with $(n{-}q)$ $\Phi$'s and $q$ $\bar{\Phi}$'s multiplet must have total U(1) charge $-2q$. This can be verified by introducing the supersymmetrized U(1) generator:\footnote{This can be derived by noting that the U(1) charge of a state is given by the difference of that in the YM copy $h_{YM}$ and that of the SYM copy $h_{SYM}$, where:
 $$h_{YM}=-\frac{1}{2}\left(\lambda_i\cdot\frac{\partial}{\partial \lambda_i}-\tilde\lambda_i\cdot\frac{\partial}{\partial \tilde\lambda_i}\right),\quad h_{SYM}=\left(1-\frac{1}{2}\eta_i\cdot\frac{\partial}{\partial\eta_i}\right)\,.$$}
 \eq\label{U1Gen}
 G_{U(1),i}=-\frac{1}{2}\left(\lambda_i\cdot\frac{\partial}{\partial \lambda_i}-\tilde\lambda_i\cdot\frac{\partial}{\partial \tilde\lambda_i}-\eta_i\cdot\frac{\partial}{\partial\eta_i}+2\right)\,.
 \eqe
One can explicitly verify that all amplitudes in eq.(\ref{N4Amps}) have the correct overall U(1) charge.

%%%%%%%%%%%%%%%%%%%%%%%%%%%%%%%%%%%%%%%%%%%%%%%%%%%%%%%%%%%%%
\subsection{Soft behaviour with anomalous U(1) symmetry }
%%%%%%%%%%%%%%%%%%%%%%%%%%%%%%%%%%%%%%%%%%%%%%%%%%%%%%%%%%%%%
 
We now consider the scalar soft limits. Possible contributions to $N_{\alpha,\beta}$ in eq.(\ref{Result}), stems from the original three-point vertices in the supergravity action as well as those that arise from the anomalous effective action. For the former, the contribution is zero since it can be derived from the tree-level soft theorems~\cite{Us}, which must vanish due to the scalar's role as parameterising the degenerate vacua. For the later, the corresponding three point vertex is $R^2\phi$ and $R^2\bar{\phi}$, which would generate a term that is of order $q^2$ in the numerator, and thus, in the $q\rightarrow0$ limit, suppresses the $1/p_i\cdot q$ propagator leading to a vanish result. Thus in the single-soft-scalar limit the amplitudes of $\mathcal{N}=4$ supergravity behave as\footnote{Here and in what follows we suppress a factor of ${i \over 4 \pi^2} ({\kappa \over 2})^{n} $ for each one-loop gravity amplitude.},
\eq\label{U1Result}
M_{\alpha,\beta,\pi}|_{q\rightarrow0}=\langle \alpha|\delta \Gamma|\beta\rangle=\frac{1}{2}\sum_{i}G_{U(1),i}M^{\Gamma}_{\alpha,\beta}\,.
\eqe 
Thus the presence of anomalies implies that the single-soft-scalar limits of $\mathcal{N}=4$ supergravity amplitudes are non-vanishing even for U(1) preserving amplitudes.  

A simple example of eq.(\ref{U1Result}) was already presented in~\cite{N4Radu}. Consider the soft scalar limit of $\overline{\rm MHV}^{(n,0)}$ obtained from tensoring the all-plus YM with the $\overline{\rm MHV}$ SYM amplitude. According to eq.(\ref{U1Result}), the single soft-limit is given by:.\footnote{We've performed a Fourier transform to convert the $\eta$ representation to that of $\tilde{\eta}$:
$$\int d^{4n}\bar{\eta}\; \bar{\eta}^qe^{\bar{\eta}\eta}\sim \eta^{4n-q}\,.$$
}
\eq
\overline{\rm MHV}^{(n{+}1,0)}:\quad \mathcal{M}_{n+1}|_{q\rightarrow0}=\left[-2-\sum_{i}\frac{1}{4}\left(\lambda_i\cdot\frac{\partial}{\partial \lambda_i}-\tilde\lambda_i\cdot\frac{\partial}{\partial \tilde\lambda_i}+\tilde{\eta}_i\cdot\frac{\partial}{\partial\tilde{\eta}_i}-2\right) \mathcal{M}_{n}\right]\,,
\eqe
Note that the above result applies to the entire super amplitude. That this is the case is because we've chose to express ${\rm MHV}^{(n,0)}$ in anti-chiral superspace $\tilde{\eta}$, where the scalar $\phi$ sits at the top of the $\Phi$ multiplet, which means that for scalar to be on leg $i$ we simply set $\tilde{\eta}_i=0$. The extra $-2$ is due to the fact that the U(1) charge of the component amplitudes is -4 of that of the super amplitude. Putting everything together, we find that 
\eq
\overline{\rm MHV}^{(n{+}1,0)}:\quad \mathcal{M}_{n+1}|_{q\rightarrow0}=(n-3) \mathcal{M}_{n}\,,
\eqe
in agreement with~\cite{N4Radu}. As another example, consider ${\rm MHV}^{(3,1)}$. Let's consider the component amplitude $\langle \bar{\phi}_1h_2^{+2}\phi_3\phi_4\rangle$. Using eq.(\ref{Super}) it is given by:
\eq\label{sample}
\langle \bar{\phi}_1h_2^{+2}\phi_3\phi_4\rangle=\frac{[23][34][42]\langle34\rangle^3}{\langle23\rangle\langle42\rangle}
\eqe
From the soft theorem in eq.(\ref{U1Result}), this would yield:
\eqa
\nonumber\langle \bar{\phi}_1h_2^{+2}\phi_3\phi_4\rangle|_{p_4\rightarrow 0}&=&\frac{1}{2}\sum_iG_{U(1),i}\langle \bar{\phi}_1h_2^{+2}\phi_3\rangle=0\\
\langle \bar{\phi}_1h_2^{+2}\phi_3\phi_4\rangle|_{p_1\rightarrow 0}&=&\frac{1}{2}\sum_iG_{U(1),i}\langle h_2^{+2}\phi_3\phi_4\rangle=0
\eqae
where the second zero is due to the fact that $\langle h_2^{+2}\phi_3\phi_4\rangle=0$. Similarly, one can also verify that ${\rm MHV}^{(5,0)}$ has vanishing single-soft scalar limits since the resulting four-point amplitude would also be zero.

Another non-trivial check of the soft theorem eq.(\ref{U1Result}), including its normalisation is the rational terms in the five-point MHV$^{(3,2)}$ amplitude. Note this is a U(1) preserving amplitude. The purely rational part of the amplitude is given by~\cite{Dunbar}:\footnote{We've adjusted the normalization by a factor of 2 from the result in~\cite{Dunbar}. This is to be consistent with our normalization as here the amplitudes are obtained via double copy which would yield a rational term for the four-point MHV amplitude as 
$$\left(\frac{t\langle12\rangle^4}{\langle12\rangle\langle23\langle\langle34\rangle\langle41\rangle}\right)^2$$
which has an overall factor of $2$ compared with eq.(4.5). }
\eq
 R[{\rm MHV}^{(3,2)}]:\;\; \mathcal{M}_5={\delta^{8}(Q)}\left(2\frac{[34][45][53]}{\langle34\rangle\langle45\rangle\langle53\rangle}+\sum_{S_{1,2},\,Z_{3,4,5}}\frac{[34]^2}{\langle 34\rangle^2}\frac{[25]\langle23\rangle\langle24\rangle}{\langle25\rangle\langle35\rangle\langle45\rangle}\right)
\eqe
where $\Phi$ multiplet is on legs $3,4,5$ ($\bar{\Phi}$ multiplet is on $1, 2$), and $R[\,*\,]$ indicates the rational part of the amplitude. In the above $S_{1,2}$ and $Z_{3,4,5}$ indicate summing over permutations of $\{1,2\}$ and cyclic permutations of $\{3,4,5\}$. Consider for example the component amplitude $\langle \bar{\phi}_1\bar{\phi}_2\phi_3\phi_4 h^{+2}_5\rangle$, which corresponds to picking up an overall factor $\langle34\rangle^4$ from $\delta^8(Q)$, and take $p_1$ to be soft, we find:
\eq\label{Dunbar1}
R[\langle \bar{\phi}_1\bar{\phi}_2\phi_3\phi_4 h^{+2}_5\rangle]\bigg|_{p_1\rightarrow0}=\langle34\rangle^3\frac{[34][45][53]}{\langle45\rangle\langle53\rangle}
\eqe
Note that while there are non-rational terms from box and bubble integrals, they do not produce pure rational functions with non-transcendental coefficients in the soft limit. Thus according to eq.(\ref{U1Result}) this should lead to $\frac{1}{2}\times 2=1$ multiplying the following component amplitude of MHV$^{(3,1)}$:
\eq
\langle \bar{\phi}_2\phi_3\phi_4 h^{+2}_5\rangle=\frac{[53][34][45]\langle34\rangle^3}{\langle53\rangle\langle45\rangle}
\eqe
It is straightforward to see from eq.(\ref{Dunbar1}) that this is indeed the case. 

Thus we see that due to the presence of anomalies, the single soft limit of non-anomalous amplitudes are also non-zero. Rather, it is proportional to the lower point amplitude with a proportionality factor given by $1/2$ times the total U(1) charge of the lower-point amplitude.

%%%%%%%%%%%%%%%%%%%%%%%%%%%%%%%%%%%%%%%%%%%%%%%%%%%%%%%%%%%%%
\section{Soft constraints on counter terms of $\mathcal{N}=4$ supergravity} \label{section:counterterm}
%%%%%%%%%%%%%%%%%%%%%%%%%%%%%%%%%%%%%%%%%%%%%%%%%%%%%%%%%%%%%
In this section, we will discuss implications of soft constraints on the counter terms in $\mathcal{N}=4$ supergravity. Explicit computation has shown that four-point amplitudes at three loops are UV finite~\cite{N41}. On the other hand, the candidate three-loop counter term, $R^4$, has been studied using extended superspace and shown to be consistent with all known symmetries including the U(1) duality symmetry~\cite{SuperSpace1}.\footnote{It has been argued that assuming the existence of off-shell superspace where all 16 supersymmetry as well as duality symmetry are linearly realised, one can rule out such counter terms~\cite{SuperSpace2, SuperSpace3}. However, the existence of such superspace yields contradictory result in the case of matter coupled supergravity~\cite{BernCounter}} In this section, we will investigate this issue from on-shell S-matrix point of view, namely we will study whether the S-matrix generated from a single insertion of $R^4$, can be completed in a way that  is consistent with both $\mathcal{N}=4$ susy and the soft theorems implied by the duality symmetry. We first note that the operator $R^4$ cannot generate R-symmetry violating amplitudes, thus the single-soft limits of matrix elements of $R^4$ should be vanishing for respecting the soft theorems. 

We will also consider four-loop counter terms. The explicit four-loop computation shows that UV divergences are present~\cite{N42}, with the corresponding counter terms are $D^2R^4$, $D^2R^3(D^2\phi)$, $D^2R^2(D^2\phi)^2$, where the latter two manifestly violate the U(1) symmetry.  Soft theorems nicely explains the universal transcendental structure of these divergences as we will show.

At four points, the supersymmetric completion of the on-shell S-matrix of $R^4$ is unique up to an overall constant, which can be expressed as, 
\eq
\mathcal{M}_4^{R^4}(\Phi_1\Phi_2 \bar{\Phi}_3\bar{\Phi}_4)=c\, \delta^{8}(Q)[12]^4\,.
\eqe
Beyond four points, possible completion of $R^4$ is constrained by the requirement that on the factorisation poles, it must factorise into a product of $R^4$ matrix elements and a supergravity amplitude. This leaves local polynomial ambiguity, which must be of the same mass dimension as $R^4$, namely $8$. Unlike in the case of $\mathcal{N}=8$ supersymmetry~\cite{ElvangSUSY}, such local polynomials are allowed for $\mathcal{N}=4$ susy. A trivial example valid for arbitrary multiplicity would be:
\eq
\langle\Phi_1\Phi_2 \bar{\Phi}_3\bar{\Phi}_4\cdots\bar{\Phi}_n\rangle=\delta^{8}(Q)[12]^4
\eqe 
However these would correspond to anomalous matrix elements as can be seen by applying eq.(\ref{U1Gen}). In other words, it would not mix with the matrix elements generated via factorisation contributions. An example for U(1) preserving local matrix element is given by:
\eqa\label{UniqueL}
\mathcal{O}_6 := \langle\Phi_1\Phi_2 \Phi_3\bar{\Phi}_4\bar{\Phi}_5\bar{\Phi}_6\rangle=\delta^{8}(Q)([12]\eta_3+[23]\eta_1+[31]\eta_2)^4\,,
\eqae
Thus there is a one parameter family of $\mathcal{N}=4$ completions of $R^4$ at six points, once the factorisation part is given. This indicates that the test of U(1) duality via soft theorems is inconclusive at six points. At the very best one can only hope to fix the coefficient in front of $\mathcal{O}_6$. This corresponds to fixing the coefficient in front of the operator $R^4\phi\bar{\phi}$, as part of the non-linear completion of $R^4$.  A non-trivial test would be whether the linear combination of $R^4$ and $R^4\phi\bar{\phi}$ is sufficient to ensure that the seven-point matrix element satisfies the U(1) soft theorem. Such an approach has already been applied to the $E_7$ symmetry of $\mathcal{N}=8$ supergravity in~\cite{ElvangSUSYCT, ElvangSUSYCTE7}, where there due to maximal supersymmetry, there are no polynomial ambiguities at the mass dimension of $R^4$, and thus a six-point soft-theorem test is already non-trivial. $R^4$ considered here for $\mathcal{N}=4$ supergravity is the analogue of $D^8R^4$ in $\mathcal{N}=8$ as studied in~\cite{ElvangSUSYCTE7}. In that case, the non-linear completion of $D^8R^4$ with $D^4R^6$ introduces two parameter families of polynomial terms.

To obtain factorisation terms we will utilize string theory amplitudes, which can be viewed as amplitudes generated from an effective action expanded in $\alpha'$. The effective action takes the following schematic form:
\eq \label{eq:lowenergy}
\mathcal{L}=R+\alpha'c_2e^{-2\tau}R^2+\alpha'^2 c_3R^3e^{-4\tau}+\alpha'^3 c_4R^4e^{-6\tau} +\cdots\,,
\eqe
where $\tau$ is the dilaton. For half-maximal supergravity, $c_3=0$ and $c_4 = 1 + 2 \zeta_3$ while $c_2={1 \over 2}$ is a pure non-transcendental number. We stress the transcendentality of these numbers here for a reason which will become clear soon. These coefficients can be obtained by computing heterotic string amplitudes. Thus by expanding the higher-point superstring amplitudes to the order $\mathcal{O}(\alpha'^3)$, we can obtain the matrix element that is generated by the non-linear expansion from $R^4$ as well as those that are generated via factorizations. For instance at six points, one finds following factorisation diagrams:
$$\includegraphics[scale=0.9]{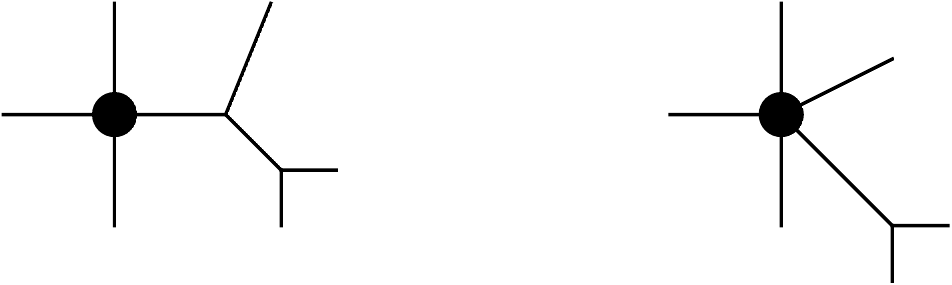}$$
The resulting matrix element is one particular completion of $R^4$, with other potential possibilities parameterised by the coefficient in front of eq.(\ref{UniqueL}). Note that due to the presence of dilaton couplings in the string effective action, one can anticipate that the single soft scalar limit of string amplitudes will not vanish. 

To compute string amplitudes with half maximal supersymmetry, one can use KLT relations with bosonic and super open string amplitudes. While open superstring amplitudes are readily available in literature~\cite{Mafra:2011nv, Mafra:2011nw}, much less is known for open bosonic string, in particular for higher-multiplicity. It turns out that for the question of our interest, we can bypass this necessity.

We first note that the $\alpha'$ expansion of string theory amplitude at $\alpha'^n$ can have at most transcendental $n$ coefficients. The coefficient of the operator $R^4$ in eq.(\ref{eq:lowenergy}) contains a $\zeta_3$, thus as we consider the $\mathcal{O}(\alpha'^3)$ term of higher-multiplicity amplitudes, we can pick out the contribution of pure $R^4$ by extracting the coefficient of $\zeta_3$. All contributions from lower dimensional operators will have lower transcendentally coefficients. We emphasise this seemingly simple observation because: {\it the leading transcendental parts of bosonic (open or close) string amplitudes in the $\alpha'$-expansion in fact precisely agree with those of superstring theories.}. At $\mathcal{O}(\alpha'^2)$ this is a well known result as the universality of $F^4$ operator in both bosonic and superstring~\cite{Stieberger:2006te}. For $\alpha'$ and $\alpha'^2$ this also holds since while the superstring do not have these terms and the bosonic string does, the coefficient for these terms are subleading transcendental. Beyond $\mathcal{O}(\alpha'^3)$, we have verified this statement for various examples of lower-point amplitudes in bosonic string amplitudes, and a general proof will be given in~\cite{alphaexpansion}. Combining with previous stated observation, namely the coefficient of $R^4$ contains $\zeta_3$, we thus conclude that the amplitudes generated by $R^4$ in half-maximal supersymmetric string theory (the leading transcendental pieces) are completely equivalent to those of maximal supersymmetry! This is the approach we will take in the following sections to compute the S-matrix generated from $R^4$ for $\mathcal{N}=4$ supergravity.

%%%%%%%%%%%%%%%%%%%%%%%%%%%%%%%%%%%%%%%%
\subsection{Counter term matrix elements from string theory amplitudes}
%%%%%%%%%%%%%%%%%%%%%%%%%%%%%%%%%%%%%%%
As we have argued, the matrix elements generated by the operator $R^4$ in half-maximal supersymmetric gravity theory can be obtained, up to the ambiguity of local terms in eq.(\ref{UniqueL}), from close string amplitudes with maximal supersymmetry. As we mentioned, one can construct close string amplitudes by the KLT relations. A general form of the KLT relation for arbitrary multiplicity is given by~\cite{BjerrumBohr:2010hn}
\bea
 {M}_n ={} &\hspace{-1.2cm}\sum_{\sigma, \gamma}
\mathcal{S}_{\alpha'}[\gamma(2,\dots,n\!-\!2)|\sigma(2,\dots,n\!-\!2)]_{k_1}
\nonumber \\
&\hspace{-0.9cm} \times
   A_n(1,\sigma(2,\dots,n\!-\! 2),n\!-\! 1,n)\,
   \tilde{A}_n(n\!-\! 1, n, \gamma(2,\dots,,n\!-\!2),1)\,,
  \label{Mnfinal}
\eea
with $\sigma, \gamma \in S_{n-3}$, and $A_n$ and $\tilde{A}_n$ are two open string amplitudes. Here the momentum kernel $\mathcal{S}$ is defined as
\begin{equation}
  \label{eq:Sn}
  \mathcal{S}_{\alpha'}[i_1,\ldots,i_k|
j_1,\ldots,j_k]_{k_i} \equiv (-1)^k (\pi\alpha')^{-k}\,
\prod_{t=1}^{k}\, \sin \big(\pi\alpha'\,(s_{i \,i_t}+ \sum_{q>t}^{k} \, \theta(i_t,i_q)\, s_{{i_t}\, {i_q} })  \big)\,,
\end{equation}
where $\theta(i_t,i_q)$ equals 1 if the ordering of
$i_t$ and $i_q$ is opposite in $\{i_1,\ldots,i_k\}$
and $\{j_1,\ldots,j_k\}$, and 0 if the ordering is the same. For our purpose we can simply take both $A_n$ and $\tilde{A}_n$ to be superstring amplitudes. A general $n$-point color-ordered open superstring gluon amplitude at tree level was worked out in~\cite{Mafra:2011nv, Mafra:2011nw}, and can be expressed in terms of $(n{-}3)!$ basis, 
\bea \label{eq:generalopen}
A(1,2, \ldots, n) = 
\sum_{\sigma \in S_{n-3}} F^{(2_{\sigma }, \ldots, (n{-}2)_{\sigma } )} 
A_{\rm SYM} (1, 2_{\sigma }, \ldots, (n{-}2)_{\sigma }, n{-}1, n) \, ,
\eea
where $A_{\rm SYM}$ is the super Yang-Mills amplitude, and the multiple hypergeometric functions are given as
\bea  \nonumber
F^{( 2, \ldots, n{-}2)} &=& ({-}1)^{n-3} \int^1_{0< z_i < z_{i+1}} \prod^{n-2}_{j=2} dz_j \left( \prod |z_{i l}|^{s_{il}} \right)
 \\
&& \times\left(  \prod^{[n/2]}_{k=2} \sum^{k-1}_{m=1} {s_{mk}  \over z_{mk}} \right)\left( \prod^{n-2}_{k=[n/2]+1} \sum^{n-1}_{m=k+1} {s_{km}  \over z_{km}}   \right) \, ,
\eea
with the Mandelstam variables defined as $s_{ij}\equiv \alpha' (k_i{+}k_j)^2$. Here we have fixed SL$(2)$ symmetry by choosing $z_1=0, z_{n-1}=1$ and $z_n = \infty$. The $\alpha'$-expansion of $F^{( 2, \ldots, n{-}2)}$ is nicely organized in terms of multi-zeta values, 
\begin{align}
F(\alpha') &= 1 + \zeta_2 P_2+\zeta_3 M_3+\zeta_2^2 P_4 + \zeta_5 M_5+\zeta_2\zeta_3 P_2M_3 + \zeta_2^3 P_6 + \frac{1}{2} \zeta_3^2 M_3 M_3 + \ldots 
 \ ,
\label{eq:alphaexpansion}
\end{align}
where the entries of the $(n-3)!\times (n-3)!$ matrices $P_w, M_w$ are degree $w$ polynomials in $s_{ij}$ with rational coefficients. Here we have shown the expansion to the order $\mathcal{O}(\alpha'^6)$, and the expressions for $P_w, M_w$ may be found in~\cite{Mafra:2011nw, Schlotterer:2012ny} and~the website~\cite{WWW}, in particular $P_2$ and $M_3$ for six and seven points are relevant for our discussion.

Before moving on to the explicit calculations, we would like to remark that the $\alpha'$-expansions of close superstring amplitudes have very nice mathematical structures as studied in~\cite{Schlotterer:2012ny, Stieberger:2013wea}. In particular, it was found that close superstring amplitudes can be expressed in a simplified KLT-like form, 
\bea
 {M}_n ={} &\hspace{-1.2cm}\sum_{\sigma, \gamma}
S [\gamma(2,\dots,n\!-\!2)|\sigma(2,\dots,n\!-\!2)]_{k_1}
\nonumber \\
&\hspace{-0.9cm} \times
   A^{\rm SYM}_n(1,\sigma(2,\dots,n\!-\! 2),n\!-\! 1,n)\,
   \tilde{A}^{\rm sv}_n(n\!-\! 1, n, \gamma(2,\dots,,n\!-\!2),1)\,,
  \label{Mnfinal2}
\eea
where $S$ now is the usual field theory KLT momentum kernel, namely $\mathcal{O}(\alpha'^0)$ order of $\mathcal{S}_{\alpha'}$ defined in eq.(\ref{eq:Sn}), thus
\begin{equation}
  \label{eq:S0}
  S[i_1,\ldots,i_k|
j_1,\ldots,j_k]_{k_i} \equiv (-1)^k \,
\prod_{t=1}^{k}\,  \big( s_{i \,i_t}+ \sum_{q>t}^{k} \, \theta(i_t,i_q)\, s_{{i_t}\, {i_q} }  \big)\,.
\end{equation}
 Furthermore, as indicated in eq.(\ref{Mnfinal2}) one amplitude entering the above formula is simply the super Yang-Mills field theory amplitude, while the other amplitude $\tilde{A}^{\rm sv}$ can be obtained from open string amplitudes in eq.(\ref{eq:generalopen}) via the so-called single-value projection\footnote{For more detailed discussion on the single-value projection, please see the reference~\cite{Schlotterer:2012ny}.}, 
\bea
{\rm sv}: \zeta_{n_1, \ldots, n_r} \mapsto \zeta^{\rm sv}_{n_1, \ldots, n_r} \, .
\eea
In particular, for our purpose to the order $\mathcal{O}( \alpha'^3)$, it projects
\bea
{\rm sv}: \zeta_2 \mapsto 0 \, , \quad \zeta_3 \mapsto  2\zeta_3 \, .
\eea
So in practice the close string formula eq.(\ref{Mnfinal2}) is a great simplification of the usual KLT formula eq.(\ref{Mnfinal}). In the following, we will use both formulas, as a double check, to study close string amplitudes in the soft limits we are interested in.  

%%%%%%%%%%%%%%%%%%%%%%%%%%%%%%%%%%%%%%%%%
\subsection{Soft constraints on $R^4$} \label{section:R4soft}
%%%%%%%%%%%%%%%%%%%%%%%%%%%%%%%%%%%%%%%%%%
We begin with six-point NMHV amplitudes, which can be obtained from the KLT relation in eq.(\ref{Mnfinal}) having both sides of the relation with NMHV open string amplitudes. Here we present one particular example: 
\eqa \nonumber
M_6(\bar{\phi}_1, h^{+2}_2, h^{+2}_3, h^{-2}_4, h^{-2}_5, \phi_6 ) \sim 
A_6(g^+_1, g^{+}_2,g^{+}_3,g^{-}_4,g^{-}_5, g^-_6 )  \times \tilde{A}_6(g^-_1, g^{+}_2,g^{+}_3,g^{-}_4,g^{-}_5, g^+_6 ) \,,
\eqae
where $\sim$ indicates we leave out kinematic factors coming from the KLT kernal and the summation over permutations. Use the KLT relation and $\alpha'$-expansion of $F^{(2_{\sigma }, 3_{\sigma }, 4_{\sigma } )}$ to the order $\mathcal{O}(\alpha'^3)$, we find that, in the single soft limit, the six-point amplitude reduces to a rather simple expression\footnote{In the appendix \ref{appendix:scalarfromgluon}, we derive the same soft-scalar limit for the heterotic string amplitude by using the known soft-gluon theorems up to subleading order. The result confirms the universality of the leading transcendental parts of amplitudes in different string theories.}
\eqa  \label{eq:6ptphiphibar}
&{}& M^{\rm string}_6(\bar{\phi}_1, h^{+2}_2, h^{+2}_3, h^{-2}_4, h^{-2}_5, \phi_6 )|_{p_6 \rightarrow 0}
\cr 
&=& 12 \zeta_3 [23]^4 \langle 45\rangle^4 
= 2 M^{\rm string}_5(\bar{\phi}_1, h^{+2}_2, h^{+2}_3, h^{-2}_4, h^{-2}_5) \, . 
\eqae
The fact that string theory yields a non-vanishing single soft limit is not surprising, since the operator $R^4$ is dressed with a dilaton factor $e^{-6\tau}$, which would generate amplitudes with non-vanishing single-soft-scalar limits. In other words, string theory yields a duality violating combination of operator $R^4$ and its non-linear completion $R^4\phi\bar\phi$. From the explicit six-point result, it is easy to see that to make string amplitudes to be consistent with the soft theorems at six points, one can simply add the local term $\mathcal{O}_6$ in eq.(\ref{UniqueL}) with right coefficient $-12 \zeta_3$, which then precisely cancels the string amplitude in eq.(\ref{eq:6ptphiphibar}) in the soft limit. Said in another way, one can add a local operator $-12 \zeta_3 R^4\phi\bar\phi$ to the low-energy effective action of string theory such that the amplitudes satisfy the soft theorem, at least at six points.

The non-trivial question is then whether string amplitudes with this additional operator $-12 \zeta_3 R^4\phi\bar\phi$ continue to satisfy the soft theorems at higher multiplicity? To test this, we consider seven-point amplitudes, for instance:
\eqa
&{}& M_7(\bar{\phi}_1, h^{-2}_2, h^{-2}_3, h^{+2}_4, h^{+2}_5, h^{+2}_6, \phi_7 ) \cr
&\sim& A_7(g^+_1, g^-_2, g^-_3, g^+_4, g^+_5, g^+_6, g^-_7 )
\times
A_7(g^-_1, g^-_2, g^-_3, g^+_4, g^+_5, g^+_6, g^+_7 ) \, .
\eqae
The single soft limit of above amplitude is non-vanishing, and yields a non-local result. An explicit numerical computation reveals that the soft limit of string amplitude $M^{\rm string}_7$ is simply proportional to the corresponding lower-point amplitude with the soft leg removed, namely, 
\eqa  \label{eq:sevenpts} \nonumber
M^{\rm string}_7(\bar{\phi}_1, h^{-2}_2, h^{-2}_3, h^{+2}_4, h^{+2}_5, h^{+2}_6, \phi_7 )\big{|}_{p_7 \rightarrow 0}
=
2M^{\rm string}_6(\bar{\phi}_1, h^{-2}_2, h^{-2}_3, h^{+2}_4, h^{+2}_5, h^{+2}_6) \, . \\
\eqae
The fact that the soft limit is non-local makes the seven-point test extremely non-trivial, since any possible additional local polynomials cannot cancel it, only the $R^4\phi\bar\phi$ operator added at six-points can do the job. Our task is then to compute the seven-point matrix element generated from the additional operator $-12 \zeta_3 R^4\phi\bar\phi$. To do so, we first note that from Feynman rules we have a simple relation between the amplitude generated from operator $R^4\phi\bar\phi$ and that from operator $R^4\phi \phi$, namely
\eqa
M(\phi,\bar\phi, \cdots)\big{|}_{R^4\phi\bar\phi} = {1 \over 2} M(\phi,\phi, \cdots)\big{|}_{R^4\phi \phi} \, ,
\eqae
here we have taken into account the symmetry factor from $\phi^2$. Furthermore, the amplitude generated from $R^4\phi \phi$ does not mix with any other operator in the string theory effective action at $\mathcal{O}(\alpha'^3)$,  and thus generates precisely the relevant amplitudes for us by the $\alpha'$ expansion. At six points we find,
\eqa \label{eq:6ptphiphi}
 M^{\rm string}_6(\phi_1, h^{+2}_2, h^{+2}_3, h^{-2}_4, h^{-2}_5, \phi_6 )
= 24 \zeta_3 [23]^4 \langle 45\rangle^4 \, ,
\eqae
which can be obtained from the KLT relation with MHV and $\overline{\rm MHV}$ open string amplitudes
\eqa \nonumber
M_6(\phi_1, h^{+2}_2, h^{+2}_3, h^{-2}_4, h^{-2}_5, \phi_6 ) \sim 
A_6(g^-_1, g^{+}_2,g^{+}_3,g^{-}_4,g^{-}_5, g^-_6 )  \times \tilde{A}_6(g^+_1, g^{+}_2,g^{+}_3,g^{-}_4,g^{-}_5, g^+_6 ) \,.
\eqae
We thus see that the six-point amplitude generated from local operator $-12 \zeta_3 R^4\phi\bar\phi$ is simply given by $-1/2$ multiplying with string amplitude with two $\phi$'s. It is easy to see that such relation should hold for general multiplicity, namely
\eqa \label{eq:relation}
M(\phi, \bar{\phi}, \ldots )\big{|}_{\rm -12 \zeta_3 R^4\phi\bar\phi}
= - {1 \over 2} M^{\rm string}_{\alpha'^3}(\phi, \phi, \ldots ) \, .
\eqae
So now we consider the following seven-point close string amplitude
\eqa
M^{\rm string}_7({\phi}_1, h^{-2}_2, h^{-2}_3, h^{+2}_4, h^{+2}_5, h^{+2}_6, \phi_7 ) \, .
\eqae
Via the KLT relations, 
\eqa
&{}& M_7(\phi_1, h^{-2}_2, h^{-2}_3, h^{+2}_4, h^{+2}_5, h^{+2}_6, \phi_7 ) \cr
&\sim& A_7(g^-_1, g^-_2, g^-_3, g^+_4, g^+_5, g^+_6, g^-_7 )
\times
A_7(g^+_1, g^-_2, g^-_3, g^+_4, g^+_5, g^+_6, g^+_7 ) \, , 
\eqae
a numerical study shows that the soft limit of this amplitude is given by
\eqa \nonumber
M^{\rm string}_7({\phi}_1, h^{-2}_2, h^{-2}_3, h^{+2}_4, h^{+2}_5, h^{+2}_6, \phi_7 )\big{|}_{p_7 \rightarrow 0}
=
4M^{\rm string}_6({\phi}_1, h^{-2}_2, h^{-2}_3, h^{+2}_4, h^{+2}_5, h^{+2}_6) \, . 
\\
\eqae
From the relation eq.(\ref{eq:relation}) and the above result, we see that seven-point amplitude generated from $-12 \zeta_3 R^4\phi\bar\phi$ would in the single-scalar soft limit, cancel that from the string amplitude in eq.(\ref{eq:sevenpts}). In other words, the non-linear completion fixed by the soft-theorems at six point satisfies the single soft constraint at seven points as well. 

To fully confirm that the operator respects the duality symmetry, one should further check the matrix elements constructed in this way respect the double-soft-scalar theorems. The double-soft-scalar theorems corresponding to duality symmetry for $\mathcal{N}=4$ supergravity were derived in~\cite{US1}, 
\eqa \label{eq:doublesoftscalar}
&{}&\bigg[M_n\left( 1,2,\cdots, n{-}2, \phi_{n-1}, \bar{\phi}_n \right)-
M_n\left( 1,2,\cdots, n{-}2, \bar{\phi}_{n-1}, \phi_n \right)
\bigg]\bigg|_{p_{n-1}, \,  p_n \rightarrow0}\\
&&=\sum_{a=1}^{n-2} \frac{p_a\cdot (p_{n-1}-p_n)}{2p_a\cdot(p_{n-1}+p_n)}\left(R_{a}\right) M_{n-2} \, ,
\eqae
where anti-symmetrization is introduced to remove the polluted soft-graviton singularity, and the U(1) R-symmetry generator is given by, 
\eqa
R_{a}=\sum_{I}\eta_a^{I}\frac{\partial}{\partial \eta_a^I} \quad (a\in \Phi )\,, \quad\quad R_{a}=\sum_{I}\eta_a^{I}\frac{\partial}{\partial \eta_a^I}-4\quad (a\in \overline{\Phi} )\,,
\eqae
where $\Phi$ and $\overline{\Phi}$ multiplets are defined in eq.(\ref{Super}). 

First we note that the local term $\mathcal{O}_6$ in eq.(\ref{UniqueL}) does not contribute in the double-soft-scalar limits due to the above anti-symmetrisation procedure. The double-soft scalar limits of $\mathcal{O}_6$ is symmetric. Thus one only needs to check whether the string amplitudes respect to the soft theorems. As we proved in section \ref{section:double-soft}, the double-soft-scalar theorems are intact even there is presence of anomaly. Indeed string amplitudes can be considered as $\alpha'$ corrections to supergravity amplitudes, which break R-symmetry in supergravity theories, we thus expect that the string amplitudes should respect the double-soft theorems. Some such examples have been observed in~\cite{Lance1, ElvangSUSYCT} for the double-soft-scalar theorems for $\mathcal{N}=8$ supergravity~\cite{Simple}. 

We have also explicitly checked that six- and seven-point close string amplitudes indeed non-trivially satisfy the double-soft-scalar theorems of $\mathcal{N}=4$ supergravity, eq.(\ref{eq:doublesoftscalar}). Let us present some examples. For instance, at six points, for the following particular helicity configuration we find, 
\eqa
M^{\rm string}_6(h^+_1, \phi_2, \bar{\phi}_3, h^-_4, \phi_5,  \bar{\phi}_6)\big{|}_{p_5\, p_6 \rightarrow 0}
&=&
 2 \left[ \left(\frac{p_2\cdot (p_5-p_6)}{p_2\cdot(p_5+p_6)} -
 \frac{p_3\cdot (p_5-p_6)}{p_3\cdot(p_5+p_6)} \right)\right. \cr
&{}&\times  \left. M^{\rm string}_4(h^+_1, \phi_2, \bar{\phi}_3, h^-_4) \right] \, .
\eqae
This is indeed in the agreement with eq.(\ref{eq:doublesoftscalar}) by noting the U(1) R-symmetry charges respect to different states are given as:
\eqa
R_a(h^+)=0 \,,\quad  R_a(h^-)=0 \, ,   \quad  R_a(\phi)=4 \, ,  \quad  R_a(\bar{\phi})=-4 \, .
\eqae
Similarly at seven points, the numerical check shows that, 
\eqa
M^{\rm string}_7(\bar{\phi}_1, \phi_2, h^+_3, h^+_4, h^-_5, \bar{\phi}_6, \phi_7)\big{|}_{p_6\, p_7 \rightarrow 0}
&=&
 2 \left[ \left(\frac{p_2\cdot (p_5-p_6)}{p_2\cdot(p_5+p_6)} -
 \frac{p_1\cdot (p_5-p_6)}{p_1\cdot(p_5+p_6)} \right)\right. \cr
&{}&\times  \left. M^{\rm string}_5(\bar{\phi}_1, \phi_2, h^+_3, h^+_4, h^-_5) \right] \, .
\eqae
From all the above examples we studied, we conclude that the low-energy effective action of string theory with the additional operator $-12 \zeta_3 R^4\phi\bar\phi$ provides us a duality invariant combination for the operator $R^4$ with respective to the single- and double-soft-scalar theorems. 

Finally we would like to remark that we have also found that the double-soft-fermion theorems of supergravity theories proposed in~\cite{US3} are satisfied by string amplitudes, although the symmetry principle behind the soft-fermion theorems is currently unclear. We take a seven-point amplitude as a very non-trivial example, for instance we find, 
\eqa 
&{}&M^{\rm string}_7(\phi^{4567}, h^+, h^+, h^-, \phi^{8124}, \psi^{123}, \psi^{35678})\big{|}_{p_6,\, p_7 \rightarrow 0} \cr
&=&
{\langle 7|5|6] \over 2 p_5 \cdot (p_6 + p_7) } M^{\rm string}_5(\phi^{4567}, h^+, h^+, h^-, \phi^{8123} ) \cr 
&{}&
-
{\langle 7|1|6] \over 2 p_1 \cdot (p_6 + p_7) } M^{\rm string}_5(\phi^{3567}, h^+, h^+, h^-, \phi^{8124} ) \, ,
\eqae
which is precisely in the form of the double-soft fermion theorems in eq.(\ref{eq:doublefermion}). Note the double-soft fermion theorems were derived from tree-level amplitudes in supergravity~\cite{US3}, here we see in examples that the theorems continue to hold for string amplitudes even though the R-symmetry of supergravity is now broken by the $\alpha'$ corrections. 

The study of compatibility of higher-dimensional operators with duality symmetry is relevant for the UV behavior the theory. Indeed the existence of available local counter terms are generally tied to the appearance of leading ultraviolet divergences in the loop expansion. Our analysis shows that $R^4$ is an acceptable counter term for $\mathcal{N}=4$ supergravity, and duality symmetry does not forbid its appearance as a counter term for UV divergence. Thus our analysis shows that supersymmetry combined with the duality symmetry are not enough for explaining the three-loop finiteness of $\mathcal{N} =4$ supergravity~\cite{N41}.

%%%%%%%%%%%%%%%%%%%%%%%%%%%%%%%%%%%%%%%%%
\subsection{Implications for four-loop counter terms}
%%%%%%%%%%%%%%%%%%%%%%%%%%%%%%%%%%%%%%%%%%
In this section, we will consider the implication of the soft theorems for the four-loop counter terms of $\mathcal{N}=4$ supergravity. It has been shown that four-point amplitudes in $\mathcal{N}=4$ supergravity are UV divergent at four loops~\cite{N42}, and the divergence is given by, 
\eq
-\frac{(1-264\zeta_3)}{144}\mathcal{T}, \quad \mathcal{T}=-st\mathcal{A}^{\rm tree}_4\left(\frac{1}{2}\mathcal{O}^{--++}+30\mathcal{O}^{++++}+\frac{3}{2}\mathcal{O}^{--++}\right)
\eqe
where $\mathcal{A}^{\rm tree}_4$ is the four-point SYM amplitude and the operators $\mathcal{O}$ correspond to distinct helicity configurations for the YM copy and are given by:
\eq
\mathcal{O}^{--++}=-4s[34]^2\langle12\rangle^2,\quad\mathcal{O}^{-+++}=12[24]^2[34]^2\langle 41\rangle^2,\quad\mathcal{O}^{++++}=-3u[34]^2[12]^2\,.
\eqe
The tensored operator correspond to $D^2R^4$, $D^2R^3(D^2\phi)$ and $D^2R^2(D^2\phi)^2$ for $\mathcal{O}^{--++}$, $\mathcal{O}^{-+++}$ and $\mathcal{O}^{++++}$ respectively. We will refer to these supergravity operators in terms of $\mathcal{O}$. A curious feature of this result is that the ratio between the leading transcendental piece $\zeta_3$ and subleading is fixed for all three distinct helicity configuration. This is surprising since the analytic structure of the amplitude in these sectors are drastically different. In particular, helicity configuration $\mathcal{O}^{-+++}$ and $\mathcal{O}^{++++}$ have vanishing unitarity cut if a tree amplitude is involved in the cut. For example at one loop, this implies that while $M_4(h_1^+h_2^+h^-_3h^-_4)$  contains logarithms, $M_4(h_1^+h_2^+\phi_3 \phi_4)$  is a pure rational function.

The universality of the aforementioned ratio can be understood from eq.(\ref{U1Result}). The presence of U(1) violating counter terms imply that there will be non-vanishing anomalous matrix elements at six-point. The U(1) soft theorem then implies that 
\eq
M^{\mathcal{O}^{--++}}_6({\phi}_1, h^{-2}_2, h^{-2}_3, h^{+2}_4, h^{+2}_5, \bar{\phi}_6 )_{p_6\rightarrow 0}=\sum_{i}G_{U(1),i}M^{\mathcal{O}^{-+++}}_5({\phi}_1, h^{-2}_2, h^{-2}_3, h^{+2}_4, h^{+2}_5)\,.
\eqe
In other words, the anomalous soft theorem requires the coefficient of $\mathcal{O}^{--++}$, which appears as an overall prefactor in the six-point amplitude, to be proportional to that of $\mathcal{O}^{-+++}$, where the proportionality constant is simply an algebraic number. If one considers the soft limit of single scalar amplitudes, then the soft-theorem will relate $\mathcal{O}^{-+++}$ to that of $\mathcal{O}^{++++}$. This explains the universality of the analytic properties for the coefficients of UV divergence in different sectors.

%%%%%%%%%%%%%%%%%%%%%%%%%%%%%%%%%%%%%%%%%%%%%%%%%%%%%
\section{Conclusions} \label{section:conclusion}
%%%%%%%%%%%%%%%%%%%%%%%%%%%%%%%%%%%%%%%%%%%%%%%%%%%%%%
In this paper, we study soft theorems of effective actions for anomalous symmetries. In particular we study the soft behaviour of amplitudes for the dilaton effective action , as well as one-loop effective action of $\mathcal{N}=4$ supergravity. We show that due to the anomaly, the current is no longer conserved, which indicates that the soft scalar limit of an $n$-point amplitude is non-vanishing and proportional to the broken generator acting on lower $(n{-}1)$-point amplitude. We show that amplitudes from dilaton effective actions satisfy the requisite soft behaviour in general, and thus soft theorems imposes no new constraint on the coefficients of the effective action, including the $a$-anomaly. For $\mathcal{N}=4$ supergravity, we show that the soft scalar limit is not only non-vanishing for U(1) non-preserving amplitudes, but for preserving ones as well, where the violation is proportional to the sum of U(1) charges of the lower point amplitude.

We extend our analysis to counter term matrix elements of $\mathcal{N}=4$ supergravity. At one and two-loops, four-dimensional pure supergravity theories are finite due to the fact that there are no susy compatible counter terms. This is no longer true at three-loops where the susy completion of $R^4$ exists. One might ask whether such an operator can be made compatible with the duality symmetry of $\mathcal{N}=4$ supergravity. This question was addressed in \cite{SuperSpace1, SuperSpace2, SuperSpace3} using superspace formalism, with different conclusions depending on the assumption for the properties of the extended superspace. Here we study the question from an on-shell point of view, thus avoiding off-shell ambiguities. In particular, we study the matrix element of $R^4$, and show that soft behaviour consistent with U(1) duality symmetry is satisfied up to seven points. We have also checked that it is consistent with the double soft scalar and fermion theorems. These result indicate that traditional symmetry arguments are insufficient to explained the three-loop finiteness of $\mathcal{N}=4$ supergravity. It would be interesting to study triple-soft limits, and see if universal soft theorems also exists. If so then one could also test whether $R^4$ satisfies such soft constraint.

In the process of deriving the matrix element, we have utilized the observation that the leading transcendental piece of $\alpha'$ expansion of string theory amplitudes are universal. This conjecture is verified by comparing the soft behaviour of the maximally supersymmetric amplitude, with that derived from string theory soft theorems for heteorotic string amplitudes. The conjecture can be proven using the analytic behaviour of the world-sheet integrals, where a detailed expose will be presented in~\cite{alphaexpansion}. Finally, the fact that soft limits relate U(1) preserving amplitudes to non-preserving ones, provides a simple explanation for the universal transcendental structure of four-loop divergences of $\mathcal{N}=4$ supergravity.

Given the soft theorems for anomalous symmetry, it would be interesting to consider bootstrapping the remaining anomalous rational amplitudes for one-loop $\mathcal{N}=4$ supergravity. Currently, only MHV$^{(n{-}2,2)}$ and MHV$^{(0,n)}$ (the conjugate of $\overline{\rm MHV}^{(n,0)}$) are known to all multiplicity. Given that MHV$^{(0,n)}$ takes an extremely simple local form, we anticipate such inverse soft construction would be applicable for MHV$^{(1,n-1)}$ as well.

%%%%%%%%%%%%%%%%%%%%%%%%%%%%%%%%%%%%%%%%%%%%%%%%%%%%%
\section{Acknowledgements}
%%%%%%%%%%%%%%%%%%%%%%%%%%%%%%%%%%%%%%%%%%%%%%%%%%%%%%

We would like to thank Wei-Ming Chen for the collaboration at an early stage of this project. We would also like to thank Massimo Bianchi, Henriette Elvang, Raffaele Marotta, Oliver Schlotterer, Stephan Stieberger and Paolo Di Vecchia for useful discussions and communications. Y-t. Huang is supported by MOST under the grant No. 103-2112-M-002-025-MY3.

\begin{appendices}
%%%%%%%%%%%%%%%%%%%%%%%%%%%%%%%%%%%%%%
\section{Higher-point dilaton amplitudes in six dimensions}\label{App}
%%%%%%%%%%%%%%%%%%%%%%%%%%%%%%%%%%%%%%%
Here we list the six and seven-point amplitude. The relevant part of the action is 
\eq
S_{eff} \;=\; \int d^6x\left\{-\frac{1}{2}(\partial \phi)^2+\mathcal{L}_2+\mathcal{L}_3\right\}+\mathcal{O}(\phi^8)
\eqe
where 
\eqa
\nonumber\mathcal{L}_2&=&\frac{b}{2f^6}\phi^2\Box^2\phi+\frac{b}{16f^8}\left[4\phi^3\Box^2\phi+\phi^2\Box^2\phi^2\right]+\frac{b}{32f^{10}}\left[5\phi^4\Box^2\phi+2\phi^3\Box^2\phi^2\right]\\
\nonumber&+&\frac{b}{128f^{12}}\left[14\phi^5\Box^2\phi+5\phi^4\Box^2\phi^2+2\phi^3\Box^2\phi^3\right]+\frac{b}{256f^{14}}\left[7\phi^5\Box^2\phi^2+5\phi^4\Box^2\phi^3\right]\\
\eqae
and 
\eqa
\nonumber\mathcal{L}_3&=&\frac{3\Delta a}{16f^8}\phi^2\Box^3\phi^2+\frac{3\Delta a}{4f^{10}}\phi^3\Box^3\phi^2+\frac{\Delta a}{48 f^{12}}\left[9\phi^4\Box^3\phi^2+4\phi^3\Box^3\phi^3\right]\\
&+&\frac{\Delta a}{40f^{14}}\left[6\phi^5\Box^3\phi^2+5\phi^4\Box^3\phi^3\right]
\eqae
Most of the above expansion was already presented in~\cite{Elvang6D}. The coefficient $b$ parameterize the ambiguity in adding Wely invariant terms to the effective action.\footnote{One can add many such term, however on-shell there is only one invariant contribution.}

From the above we can obtain the six and seven-point amplitudes with four and six derivatives. They are:
\eqa
A_6^{p^{4}}=\frac{3b}{f^{12}}\sum_{1\leq i<j\leq 6} s^2_{ij},\quad A_7^{p^{4}}=\frac{15b}{f^{14}}\sum_{1\leq i<j\leq 7} s^2_{ij}
\eqae
For $p^{6}$ the result is more complicated due to the presence of factorization poles. Nevertheless they are given by:
\eqa
\nonumber A_6^{p^{6}}&=&\left[\frac{3}{2}\Delta a -\frac{b^2}{f^4}\right]\frac{6}{f^{12}}\sum_{1\leq i<j\leq 6}s^3_{ij}+\left[\Delta a -\frac{5b^2}{8f^4}\right]\frac{3}{f^{12}}\sum_{1\leq i<j<k\leq 6}s^3_{ijk}\\
\nonumber&-&\frac{b^2}{4f^4}\frac{1}{f^{12}}\left((s^2_{12}+s^2_{23}+s^2_{13})\frac{1}{s_{123}}(s^2_{45}+s^2_{56}+s^2_{46})+{\rm perm}\right)\\
\nonumber A_7^{p^{6}}&=&\left[12\Delta a-\frac{35b^2}{8f^4}\right]\frac{3}{f^{14}}\sum_{1\leq i<j\leq 7}s^3_{ij}+\left[2\Delta a-\frac{5b^2}{8f^4}\right]\frac{9}{f^{14}}\sum_{1\leq i<j<k\leq 7}s^3_{ijk}\\
\nonumber&-&\frac{b^2}{f^{18}}\left[\frac{15}{8}s_{ij}(s^2_{kl}+s^2_{klmn})+\frac{9}{8}s_{ij}s^2_{klm}+\frac{9}{8}s_{ijk}(s^2_{mn}+s^2_{lmn})+\frac{15}{8}s_{ijkl}s^2_{mn}+{\rm perm}\right]\\
&-&\frac{b^2}{f^{18}}\frac{3}{8}\left[\frac{s^2_{ij}s^2_{lmn}}{s_{ijk}}+\frac{s^2_{ij}s^2_{mn}}{s_{ijk}}+{\rm perm}\right]
\eqae 
where in the above, permutation indicates that one sums over all distinct factorization channels.

%%%%%%%%%%%%%%%%%%%%%%%%%%%%%%%%%%%%%%
\section{Absence of five-point amplitude for vector multiplet}\label{Vector}
%%%%%%%%%%%%%%%%%%%%%%%%%%%%%%%%%%%%%%%
Theories with vectors cannot be conformal invariant in the IR, since free vectors have non-vanishing trace for the stress tensor, and hence the identification of the $a$ function becomes problematic. In any case, one can still check whether five-point amplitudes vanish due to supersymmetry. In such case the super amplitudes must carry the chiral little group $\dot{a}$ for each leg. The corresponding ansatz is 
\eqa
\mathcal{A}_5
&=&
\delta^4(Q)\bigg(\alpha\langle q_1|3_{\dot{c}}][1_{\dot{a}}|p_3|2_{\dot{b}}][4_{\dot{d}}|p_1|5_{\dot{e}}]+\beta\langle q_1|3_{\dot{c}}][1_{\dot{a}}|p_4|2_{\dot{b}}][4_{\dot{d}}|p_1|5_{\dot{e}}]
\cr
&+&
\gamma\langle q_1|3_{\dot{c}}][1_{\dot{a}}|p_4|2_{\dot{b}}][4_{\dot{d}}|p_3|5_{\dot{e}}]+\delta\langle q_1|3_{\dot{c}}][1_{\dot{a}}|p_3|2_{\dot{b}}][4_{\dot{d}}|p_3|5_{\dot{e}}]\bigg)\cr
&+&
(-)\, {\rm perm} (1,2,3,4,5)\, .
\eqae 
Subjecting the above result to the constraint $\tilde{Q}^A\mathcal{A}_5=0$, one again finds that there are no solution to the ansatz.

%%%%%%%%%%%%%%%%%%%%%%%%%%%%%%%%%%%%%%
\section{Soft scalars from soft gluons}  \label{appendix:scalarfromgluon}
%%%%%%%%%%%%%%%%%%%%%%%%%%%%%%%%%%%%%%

As the scalars of $\mathcal{N}=4$ supergravity can be obtained from the product of positive helicity and negative helicity gluons, the KLT representation allows us to utilize known soft-gluon theorems to obtain the soft-scalar limits. It is known that for the single-soft-gluon limit, open string amplitudes satisfy the same leading and subleading soft behaviours as the tree-level amplitudes in Yang-Mills theories~\cite{Schwab:2014fia, US2}.\footnote{Double-soft limits in string amplitudes were recently studied in~\cite{Volovich:2015yoa, DiVecchia:2015bfa}} In other words one can extract the soft-scalar behaviour of $R^4$ operators from the known soft-gluon limits of open string amplitudes. Here we will consider six-point amplitudes. For the convenience, we quote six-point KLT formula here, 
\eqa\label{KLTList}
\nonumber M_6&=&-%\pi \kappa^4 
(\pi \alpha' )^{-3} A(1,2,3,4,5,6) \sin(\alpha'\pi s_{12} ) \sin(\alpha'\pi s_{45} )\left[\tilde{A}(2,3,1,5,4,6)\sin(\alpha'\pi s_{13} ) \right. \\
&&+ \left. \tilde{A}(3,2,1,5,4,6)\sin(\alpha'\pi  (s_{13}+s_{23}) ) \right] + {\rm Perm} (2,3,4) \, ,
\eqae
where $A$ and $\tilde{A}$ are bosonic and super open string amplitudes, respectively. 
The soft-scalar limit of $M_6$ can be obtained by the soft-gluon limit open-string amplitudes entering the above KLT formula. Without losing generality, we will take leg $1$ to be soft, with $p_1 \rightarrow \delta p_1$. Due to the factor $\sin(\alpha'\pi s_{12} ) \sim \delta $ in the KLT relation, and the fact that the leading soft-gluon behaviour is $\delta^{-1}$, we need gluon amplitudes to the order $\mathcal{O}(\delta^0)$, which is precisely the order that amplitudes behave universally, and are given by~\cite{US2} 
\bea \label{BCFWYM}
&& A_{n}(  \{ \sqrt{\delta} \lambda_1, \sqrt{\delta} \tilde{\lambda}_1 \}^+ \, , \{  \lambda_2, \tilde{\lambda}_2 \}, \ldots \,,  
\{  \lambda_n, \tilde{\lambda}_n \})
\cr
&=& \sum_{k=0,1} \frac{1}{\delta^{1-k}}S_{\rm YM}^{(k)} (n\,1\,2)
A_{n-1}(  \{  \lambda_2, \tilde{\lambda}_2 \}\, , \ldots \,,  
\{  \lambda_n, \tilde{\lambda}_n \}) +\mathcal{O}(\delta)
\eea
where the superscript $+$ indicates the soft leg being a positive helicity gluon, and the soft factor is a differential operator at the subleading order, and is given by
\bea \label{SkYM}
S^{(k)}_{\rm YM} (n \, 1\, 2)= 
{1 \over k!} { \langle n 2 \rangle  \over \langle n 1\rangle \langle 1 2\rangle  } 
\left( { \langle 1  n \rangle \over \langle 2 n\rangle } \tilde{\lambda}_1 \cdot { \partial \over \partial
 \tilde{\lambda}_2 } +
 { \langle 1  2 \rangle \over \langle n  2\rangle } \tilde{\lambda}_1 \cdot { \partial \over \partial
 \tilde{\lambda}_n }   \right)^k \, .
\eea
The soft factor of a negative gluon can be obtained by the parity conjugate of $S^{(k)}_{\rm YM}$. To obtain the soft-scalar limit for six-point amplitudes, we thus only need five-point open string amplitudes (with above soft factors acting on them), which greatly simplifies the computation. By substituting the soft-gluon theorems eq.(\ref{BCFWYM}) (with five-point bosonic and super open string amplitudes) into the six-point KLT formula eq.(\ref{KLTList}), expand to order $\mathcal{O}(\alpha'^3)$, we numerically verified that the leading transcendental parts (namely the $\zeta_3$ part) of the single-soft scalar limits obtained in this way indeed agree with that in section \ref{section:R4soft}. 

\end{appendices}

%%%%%%%%%%%%%%%%%%%%%%%%%%%%%%%%%%%%%%%%%%%%%%%%%%%%%%%%%%%%%%%%
%%%%%%%%%%%%%%%%%%%%%%%%%%%%%%%%%%%%%%%%%%%%%%%%%%%%%%%%%%%%%%%%

\end{large}

\end{document}